\documentclass[11pt,a4paper]{article}
\pdfoutput=1
\usepackage{jheppub}

\usepackage[utf8]{inputenc}
\usepackage{amsmath}
\usepackage{amsthm}
\usepackage{tabularx}
\usepackage{tikz}
\usepackage{multirow}
\usepackage{longtable}
\usepackage{multirow}
\usepackage{amsmath}
\usepackage{amssymb}
\usepackage{mathtools}
\usepackage{flafter}
\usepackage{chngpage}
\usepackage{adjustbox}
\usetikzlibrary{shapes,arrows}
\usepackage[english]{babel}
\newcolumntype{P}[1]{>{\centering\arraybackslash}p{#1}}
\usepackage{tabularx}

\usepackage{color}
\usepackage{subcaption}

\tikzset{gauge1/.style={draw=none,minimum size=0.35cm,fill=blue,circle, draw}}
\tikzset{gauge2/.style={draw=none,minimum size=0.35cm,fill=red,circle, draw}}
\tikzset{flavour1/.style={draw=none,minimum size=0.35cm,fill=blue, regular polygon,regular polygon sides=4,draw}}
\tikzset{flavour2/.style={draw=none,minimum size=0.35cm,fill=red, regular polygon,regular polygon sides=4,draw}}

\newcommand{\G}{\mathcal{G}}
\newcommand{\m}{\mathbf{m}}
\newcommand{\AO}{\bar A (\mathcal O_\lambda)}

\preprint{Imperial/TP/17/AH/03}

\title{Nilpotent orbits and the Coulomb branch of $T^\sigma (G)$ theories: special orthogonal vs orthogonal gauge group factors}
\author[a]{Santiago Cabrera}
\author[a]{Amihay Hanany}
\author[a]{and Zhenghao Zhong}
\affiliation[a]{Theoretical Physics, The Blackett Laboratory\\
Imperial College London\\ SW7 2AZ United Kingdom}
\emailAdd{santiago.cabrera13@imperial.ac.uk}
\emailAdd{a.hanany@imperial.ac.uk}
\emailAdd{zhenghao.zhong14@imperial.ac.uk}

 \abstract{Coulomb branches of a set of $3d\ \mathcal{N}=4$ supersymmetric gauge theories are closures of  nilpotent orbits of the algebra $\mathfrak{so}(n)$. From the point of view of string theory, these quantum field theories can be understood as effective gauge theories describing the low energy dynamics of a brane configuration with the presence of orientifold planes \cite{FH00}.  The presence of the orientifold planes raises the question to whether the orthogonal factors of a the gauge group are indeed \emph{orthogonal} $O(N)$ or \emph{special orthogonal} $SO(N)$. In order to investigate this problem, we compute the Hilbert series for the Coulomb branch of $T^\sigma(SO(n)^\vee)$ theories, utilizing the \emph{monopole formula}. The results for all nilpotent orbits from $\mathfrak {so} (3)$ to $\mathfrak{so}(10)$ which are \emph{special} and \emph{normal} are presented. A new relationship between the choice of $SO/O(N)$ factors in the gauge group and the \emph{Lusztig's Canonical Quotient} $\AO$ of the corresponding nilpotent orbit is observed. We also provide a new way of projecting several magnetic lattices of different $SO(N)$ gauge group factors by the diagonal action of a $\mathbb Z_2$ group.}


\begin{document}

\maketitle
\flushbottom

\section{Introduction}

In \cite{CHMZ14} $3d$ $\mathcal N=4$ quiver gauge theories known as $T_\rho^\sigma (G)$ theories \cite{FH00,GW09} were studied for $G$ a classical Lie group of type $A$, $B$, $C$ and $D$. The Hilbert series\footnote{See \cite{C17} for a review of Hilbert series and moduli spaces of $3d\ \mathcal N=4$ theories.} for the Coulomb branch and for the Higgs branch of these theories were computed for the case when either $\sigma$ or $\rho$ are trivial partitions. For $T^\sigma(G)$ on the $B$, $C$ and $D$ cases (where $\rho$ is trivial) an interesting phenomenon occurs: there are several possible quivers for a single choice of partition $\sigma$ and the Coulomb branches of these quivers differ by a projection by a discrete group, like for example $\mathbb Z_2$.

On the other hand, Coulomb branches of $T^\sigma (G)$ theories are predicted to be closures of nilpotent orbits of the algebra $Lie(G^\vee)$ \cite{KP79,KS96,GW09,CDT13,BTX10}, where $G^\vee$ is the Langlands or GNO dual group of $G$ \cite{GNO76}. Therefore the question arises as to which of the different quivers obtained for a single theory has the closure of the nilpotent orbit as its Coulomb branch, rather than a discrete cover of the same.

The present note constitutes a systematic approach to this problem. The Hilbert series of the Coulomb branches of all the candidate quivers are computed utilizing the \emph{monopole formula} \cite{CHZ13,HM16,C17}. Analyzing the Hilbert series of each moduli space one can discern which Coulomb branch corresponds to the closure of a nilpotent orbit. The following pages contain both the quivers that resulted to have the closure of the nilpotent orbit as their Coulomb branch and the corresponding Hilbert series.

Section \ref{method} of the paper reviews the construction of $T^{\sigma}(G^\vee)$ \emph{orthosymplectic quivers} given in \cite{CHMZ14}. Section \ref{math} overviews the \emph{monopole formula}, used to compute the Hilbert series of the Coulomb branch of the quivers.  Section \ref{result} tabulates the $T^{\sigma}(G^\vee)$ quivers with the choice of $SO(N)/O(N)$ gauge nodes such that their Coulomb branch is the closure of the corresponding nilpotent orbit. Results for $\mathfrak{so}(3)$  to $\mathfrak{so}(10)$ are presented along with their Hilbert series and plethystic logarithms. Section \ref{sec:analysis} contains an analysis of the results and provides an answer to the problem. In section \ref{sec:Z2actions} a new method to project several magnetic lattices by the diagonal action of a $\mathbb Z_2$ group is introduced. This mechanism is a key part of the answer to the problem and opens the door to a new range of computations that can be performed utilizing the \emph{monopole formula}.

\section{Constructing quiver diagrams for $T^\sigma(G)$ theories}\label{method}

This section reviews the method of \cite{CHMZ14} to obtain the quiver of the $3d$ $\mathcal N=4$ superconformal field theory known as $T^\sigma (G)$. This quiver is determined by a partition $\sigma$ and a classical Lie group $G$. Please see \cite{C17} for a review on $3d$ $\mathcal N=4$ quiver gauge theories and their moduli spaces.

Two cases are considered here: $T^\sigma(SO(2n))$ and $T^\sigma(USp(2n))$. 

\subsection{$T^\sigma(SO(2n))$ quiver}

For $G=SO(2n)$ one considers partitions $\sigma\in \mathcal P_D(2n)$. $\mathcal P_D(2n)$ is the set of all partitions of the integer number $2n$ such that all even parts occur with even multiplicity. For example $\sigma=[4,4,4,4,3,2,2,1]$ is a partition of $\mathcal P_D(24)$. 

We only consider partitions such that

\begin{align}\label{eq:special}
	(d_{LS})^2(\sigma)=\sigma
\end{align}

where $d_{LS}$ is the Lusztig-Spaltenstein map \cite{Sp82} that transposes and D-collapses the partition, see the discussion in\footnote{In \cite{CM93} the Lusztig-Spaltenstein map is simply called \emph{Spaltenstein map}.} \cite{CM93}. Partitions $\sigma\in \mathcal P_D(2n)$ that satisfy equation (\ref{eq:special}) are called \emph{special}.

The orthosymplectic quiver has a chain with a number of $2n-1$ gauge nodes that alternate between the groups $SO(N_j)$ and $USp(N_j)$, starting with $USp(N_1)$ (we label the nodes from right to left, i.e. $USp(N_1)$ is the rightmost node). Note that in principle $O(N_j)$ groups can be chosen instead of $SO(N_j)$. The parameter $N_j$ can be obtained from the partition data:
\begin{align}\label{eq:gauge}
N_j = \Bigg[\sum_{k=j+1}^{l'}\rho_k\Bigg]_{+,-} -   \Bigg(\sum_{i=j+1}^{l}\sigma^T_i\Bigg), 
\end{align}

where $\sigma^T=[\sigma^T_1,\sigma^T_2,\dots,\sigma^T_l]$ is the transpose partition of $\sigma$ and $\rho=[1,1,\dots,1]$ is the trivial partition of $2n$. The symbol $[p]_+$ means the smallest even integer that is larger or equal to $p$ and $[p]_-$ is the largest even integer smaller or equal to $p$. Also $l'=2n$ is the length of $\rho$ and $l$ is the length of $\sigma^T$. ``$+$" needs to be used when computing $SO(N_j)$ gauge nodes whereas ``$-$" is for $USp(N_j)$ gauge nodes. 

A single flavor node can be attached to each gauge node. The flavor nodes also alternate: $SO(M_j)$ are attached to symplectic gauge nodes and $USp(M_j)$ are attached to orthogonal gauge nodes. The parameters $M_j$ are also determined by the partition data:

\begin{align}\label{eq:flavors}
	M_j=\sigma^T_j-\sigma^T_{j+1}
\end{align}

where zero parts $\sigma^T_{l+1}=0$, $\sigma^T_{l+2}=0$, etc. can be added at the end of partition $\sigma^T$ if necessary.

\subsection{$T^\sigma(USp(2n))$ quiver}

For $G=USp(2n)$ we consider partitions $\sigma\in \mathcal P_C(2n)$. $\mathcal P_C(2n)$ is the set of all partitions of the integer number $2n$ such that all odd parts occur with even multiplicity. For example $\sigma=[4,4,4,3,3,2,2,1,1]$ is a partition of $\mathcal P_C(24)$. Again, only special partitions such that $(d_{LS})^2(\sigma)=\sigma$ are considered.

The orthosymplectic quiver has now a chain with a number of $2n$ gauge nodes that alternates between the groups $SO(N_j)$ and $USp(N_j)$, starting with $SO(N_1)$. Note that in principle $O(N_j)$ groups can be chosen instead of $SO(N_j)$. The parameter $N_j$ can be obtained from the partition data with equation (\ref{eq:gauge}) (with the difference that in this case $\rho=[1,1,\dots,1]$ is the trivial partition of $2n+1$).

A single flavor node can be attached to each gauge node. The flavor nodes also alternate: $SO(M_j)$ are attached to symplectic gauge nodes and $USp(M_j)$ are attached to orthogonal gauge nodes. The parameters $M_j$ are also determined by the partition data as indicated in equation (\ref{eq:flavors}).

\subsection{Coulomb branches}

\cite{BTX10,CDT13} established that that the Coulomb branch of the $T^\sigma(G)$ theory should be the closure of the nilpotent orbit\footnote{For an introduction to nilpotent orbits see \cite{CH16}, we follow the same notation. A standard book on the topic is \cite{CM93}.} $\bar{ \mathcal O}_\lambda \subset \mathfrak{g}^\vee$ with $\mathfrak{g}^\vee=Lie(G^\vee)$ and

\begin{align}
	\lambda = d_{BV} (\sigma)
\end{align}

The map $d_{BV}$ is the Barbasch-Vogan map \cite{BV85} defined by equation\footnote{Note that the name \emph{Spaltenstein map} is adopted in \cite{CDT13} for the Barbasch-Vogan map. Also note that there is a mistake in \cite{CHMZ14} where the map is defined by equation (6.2): for a partition of C-type the map is defined as transposition, addition of a new part and B-collapsing; the correct order is addition of a new part, then transposition and then B-collapsing.} (2.8) of \cite{CDT13}, which is taken from equation (5) in \cite{A02}. This is summarized in table \ref{tab:BV}, where $\sigma^T$ is the transpose partition, $\sigma^\pm$ is defined as $\sigma^+:=[\sigma_1,\sigma_2,\dots,\sigma_s,1]$ and $\sigma^-:=[\sigma_1,\sigma_2,\dots,\sigma_s-1]$. $\sigma_X$, with $X=B,C,D$, is the $X$-collapse of partition $\sigma$ as defined in \cite{CM93}. $\mathcal{P}_A(n)$ is the set of all partitions of $n$; $\mathcal P_B(2n+1)$ is the set of partitions of $2n+1$ where even parts have even multiplicity; $\mathcal P_C(2n)$ is the set of partitions of $2n$ where odd parts have even multiplicity and $\mathcal P_D(2n)$ is the set of partitions of $2n$ where odd parts have even multiplicity.

\begin{table}[t]
	\centering
    \begin{tabular}{|l|l|l|}
    \hline
    Set $\mathcal P$ & $d_{BV}(\sigma)$ & Set $\mathcal P^\vee$\\ \hline
    $\mathcal {P}_A(n)$ & $\sigma^T$ & $\mathcal {P}_A(n)$\\ 
    $\mathcal {P}_B(2n+1)$ & $(\sigma^T)^-_{\ \ C}$ & $\mathcal {P}_C(2n)$\\ 
    $\mathcal {P}_C(2n)$ & ($\sigma^+)^T_{\ \ B}$ & $\mathcal {P}_B(2n+1)$\\ 
    $\mathcal {P}_D(2n)$ & $(\sigma^T)_D$ & $\mathcal {P}_D(2n)$\\ \hline
    
    \end{tabular}
    \caption{Barbasch-Vogan map $d_{BV}$. The first column is the set of partitions $\mathcal P$ such that $\sigma\in \mathcal P$. The last column is the set of partitions such that $d_{BV}(\sigma)\in \mathcal P^\vee$.}
    \label{tab:BV}
\end{table}

This means that for $G=SO(2n)$ the Coulomb branches of $T^\sigma(SO(2n))$ for special partitions $\sigma$ cover the whole set of closures of special nilpotent orbits of $\mathfrak{so}(2n)$. For $G=USp(2n)$ Coulomb branches of $T^\sigma (USp(2n))$ for special partitions $\sigma$ cover the whole set of closures of special nilpotent orbits of $\mathfrak{so}(2n+1)$. There are some exceptions for this statement: closures of nilpotent orbits that are non-normal have not been found as Coulomb branches of any theory\footnote{It remains a challenge to figure out if Coulomb branches can be \emph{non-normal}, as their Higgs branch counterparts. The challenge has recently gained importance since Nakajima's mathematical construction for the Coulomb branch \cite{N15} has been shown to always be a normal variety \cite{BFN16}.}.

\subsection{Example: $\lambda=[3,3,1]$ orbit of $\mathfrak{so}(7)$}
Let us look at an example for the closure of the nilpotent orbit $\bar{\mathcal O}_{[3,3,1]}\subset \mathfrak{so}(7)$. This means that $\lambda=[3,3,1]$ and $\mathfrak{g}^\vee=\mathfrak{so}(7)$. Therefore the relevant quiver is $T^\sigma(G)$ with:

\begin{align}
	\begin{aligned}
		G&=SO(7)^\vee = USp(6)\\
		\sigma&=d_{BV}[3,3,1]=[2,2,2]\\
	\end{aligned}
\end{align}

Note that $\lambda=d_{BV}(\sigma)$ implies $\sigma=d_{BV}(\lambda)$ because we are restricting $\sigma$ to the set of \emph{special} partitions. The quiver theory is $T^{[2,2,2]}(USp(6))$. The necessary partition data in order to obtain the quiver is:

\begin{align}
	\begin{aligned}
		\sigma^T&=[2,2,2]^T=[3,3]\\
		\rho&=[1,1,1,1,1,1,1]\\
	\end{aligned}
\end{align}

The $N_j$ parameters for the gauge nodes are: 

\begin{align}
	\begin{aligned}
		N_1 &= [6]_+ -3 = 3\\
		N_2 &= [5]_- = 4\\
		N_3 &= [4]_+ = 4\\
		N_4 &= [3]_- = 2\\
		N_5 &= [2]_+ = 2\\
		N_6 &= [1]_-= 0\\
	\end{aligned}
\end{align}

The $M_j$ parameters for the flavor nodes are:

\begin{align}
	\begin{aligned}
		M_1 &= 3-3=0\\
		M_2 &= 3-0=3\\
		M_3 &=0-0=0\\
		M_4 &= 0-0=0\\
		M_5 &=0-0=0\\
		M_6 &=0-0=0\\
	\end{aligned}
\end{align}

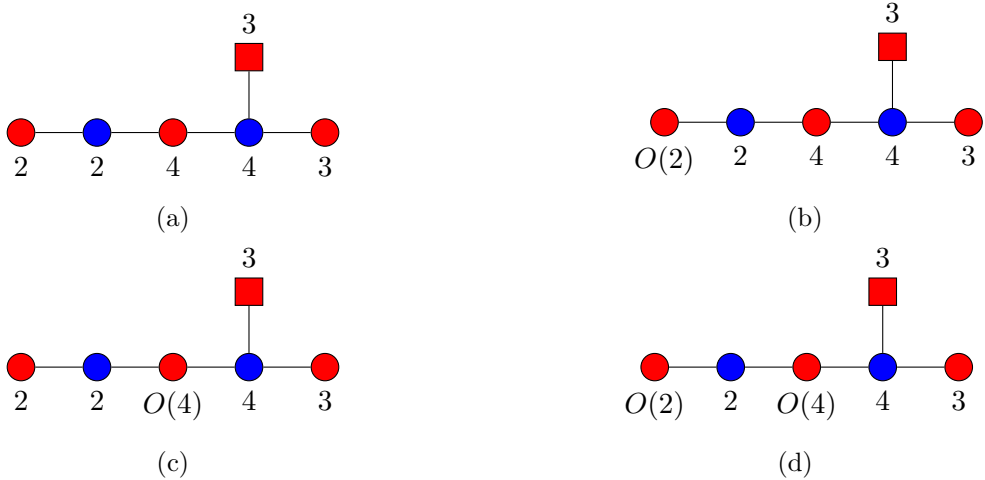
\begin{figure}[t]
	\centering
	\begin{subfigure}[t]{.45\textwidth}
		\centering
		\begin{tikzpicture}
			\node (g1) [gauge2,label=below:{$2$}] {};
			\node (g2) [gauge1,right of=g1,label=below:{$2$}] {};
			\node (g3) [gauge2,right of=g2,label=below:{$4$}] {};
			\node (g4) [gauge1,right of=g3,label=below:{$4$}] {};
			\node (g5) [gauge2,right of=g4,label=below:{$3$}] {};
			\node (f1) [flavour2,above of=g4,label=above:{$3$}] {};
			\draw (g4)--(f1) 
				(g1)--(g2) (g2)--(g3) (g3)--(g4) (g4)--(g5);
		\end{tikzpicture}
		\caption{}
	\end{subfigure}
	\hfill
	\begin{subfigure}[t]{.45\textwidth}
		\centering
		\begin{tikzpicture}
			\node (g1) [gauge2,label=below:{$O(2)$}] {};
			\node (g2) [gauge1,right of=g1,label=below:{$2$}] {};
			\node (g3) [gauge2,right of=g2,label=below:{$4$}] {};
			\node (g4) [gauge1,right of=g3,label=below:{$4$}] {};
			\node (g5) [gauge2,right of=g4,label=below:{$3$}] {};
			\node (f1) [flavour2,above of=g4,label=above:{$3$}] {};
			\draw (g4)--(f1) 
				(g1)--(g2) (g2)--(g3) (g3)--(g4) (g4)--(g5);
		\end{tikzpicture}
		\caption{}
	\end{subfigure}
	\hfill
	\begin{subfigure}[t]{.45\textwidth}
		\centering
		\begin{tikzpicture}
			\node (g1) [gauge2,label=below:{$2$}] {};
			\node (g2) [gauge1,right of=g1,label=below:{$2$}] {};
			\node (g3) [gauge2,right of=g2,label=below:{$O(4)$}] {};
			\node (g4) [gauge1,right of=g3,label=below:{$4$}] {};
			\node (g5) [gauge2,right of=g4,label=below:{$3$}] {};
			\node (f1) [flavour2,above of=g4,label=above:{$3$}] {};
			\draw (g4)--(f1) 
				(g1)--(g2) (g2)--(g3) (g3)--(g4) (g4)--(g5);
		\end{tikzpicture}
		\caption{}
	\end{subfigure}
	\hfill
	\begin{subfigure}[t]{.45\textwidth}
		\centering
		\begin{tikzpicture}
			\node (g1) [gauge2,label=below:{$O(2)$}] {};
			\node (g2) [gauge1,right of=g1,label=below:{$2$}] {};
			\node (g3) [gauge2,right of=g2,label=below:{$O(4)$}] {};
			\node (g4) [gauge1,right of=g3,label=below:{$4$}] {};
			\node (g5) [gauge2,right of=g4,label=below:{$3$}] {};
			\node (f1) [flavour2,above of=g4,label=above:{$3$}] {};
			\draw (g4)--(f1) 
				(g1)--(g2) (g2)--(g3) (g3)--(g4) (g4)--(g5);
		\end{tikzpicture}
		\caption{}
	\end{subfigure}
	\hfill
	\caption{Candidate quivers for $T^{[2,2,2]}(USp(6))$ theory. Gauge nodes are denoted by a circle and flavor nodes are represented by a square. Red gauge nodes with label $N_j$ denote $SO(N_j)$ groups and blue gauge nodes with label $N_j$ denote $USp(N_j)$ gauge groups. The red square node with label $M_2=3$ denotes an $SO(3)$ flavor node. When the gauge group $O(N_j)$ is chosen instead of $SO(N_j)$ the group is written explicitly.}
	\label{fig:example}
\end{figure}

This gives four possible quivers with different Coulomb branch $\mathcal C$ depending on the choice $SO/O(N_j)$ for $N_j$ even\footnote{Note that only orthogonal groups with $O(N_j)$ with $N_j$ even are considered. This is a straight forward exercise since a description of their magnetic lattices and dressing factors has been provided in \cite{CHMZ14}. The analysis of $N_j$ odd is left for future work.}, depicted in figure \ref{fig:example}. The aim of this paper is to compute the Coulomb branch of the four quivers in order to determine which of them has the closure of the nilpotent orbit $\mathcal C=\bar{\mathcal O}_{[3,3,1]}$ as its Coulomb branch, and therefore corresponds to $T^{[2,2,2]}(USp(6))$. In this case the answer is the quiver in figure \ref{fig:example}(a), as is shown in the third row of table \ref{table5}.

\section{The monopole formula}\label{math}
\subsection{The Hilbert series}

A Hilbert series $H(t)$ can be computed in order to determine the Coulomb branch of each quiver, see \cite{C17}. In this case the Hilbert series can be obtained via the \emph{monopole formula} introduced in \cite{CHZ13}, which counts dressed monopole operators in the chiral ring. We shall only consider the unrefined Hilbert series $H(t)$, given as:
\begin{equation}
H(t)= \sum_{\m \in \Gamma^*_{\G^\vee}/W_{\G^\vee}} t^{\Delta (m)}P_\G(\m ;t),
\label{2}
\end{equation}
where $\G$ is the gauge group and the magnetic fluxes $\m$ are summed over the dominant Weyl chamber of the weight lattice of $\G^\vee$, the GNO dual group of $\G$ \cite{GNO76}. $\m$ breaks $\G$ into a residual gauge group $\G_\m$. In $3d\ \mathcal{N}=4$, the dressing factors $P_\G(\m;t)$ are given by the dimensions $d_i$ of the Casimir invariants of $\G_\m$:

\begin{align}
	P_\G(\m;t) =\prod_{i=1}^{r} \frac{1}{(1-t^{d_i(\G_\m)})}
\end{align}

where $r=rank(\G)$. $\Delta(\m)$ is the \emph{conformal dimension} of the bare monopole operators: 

\begin{align}
	\Delta (\m) = - \sum_{\alpha \in \Delta_{+}}|\alpha(\m)|+ \frac{1}{2}\sum_{i=1}^{n}\sum_{\rho _i \in \mathcal{R}_i}|\rho_i (\m)|,
\end{align}

where the first sum is over positive roots of the gauge group $\G$ and it is the contribution from the vector multiplets of the theory. The second sum is over weights of the representation $\mathcal{R}_i$ under which the hypermultiplets of the theory transform (do not confuse the symbol $\rho_i$ for the weights with the previous symbol $\rho$ for the trivial partitions).  


\subsection{The plethystic logarithm}
Once a Hilbert series $H(t)$ is obtained, it is useful to analyze its plethystic logarithm \cite{FHH07}, $PL(H(t))$ defined as:

\begin{equation}
PL(H(t)) = \sum_{k=1}^{\infty}\frac{\mu (k)}{k}\log(H(t^k))
\end{equation}
where $\mu (k)$ is the M\"obius function. The $PL(H(t))$ is a generating function that counts generators of the chiral ring and their relations. Since a moduli space that is the closure of a nilpotent orbit restricts the dimension of the generators of the chiral ring to $\Delta=1$ (see discussion in sections 3.1 and 3.2 of \cite{CH16}) the $PL(H(t))$ provides invaluable information for the present analysis. 

\section{Quivers for closures of nilpotent orbits}\label{result}

This section contains the main results of the paper. The quiver diagrams of theories $T^\sigma(G)$ with the right choice of $SO/O(N_j)$ gauge nodes so their Coulomb branch $\mathcal C$ is the closure of nilpotent orbit $\bar{\mathcal O}_\lambda\subset\mathfrak{so}(n)$ are tabulated for $3\leq n \leq 10$. Remember that $\sigma=d_{BV}(\lambda)$ and vice versa. The Hilbert series and plethystic logarithms of the Coulomb branch of these quivers are also tabulated. 

\subsection{Tables: quivers}\label{sec:quivers}
For the tables in this section:
\begin{itemize}
	\item All red nodes with label $N$ denote $SO(N)$ groups and all blue nodes with label $N$ denote $USp(N)$ groups. Red nodes corresponding to $O(N)$ groups are explicitly labeled as $O(N)$.
	\item A single quiver is provided for each closure of a special nilpotent orbit $\bar{\mathcal O}_\lambda\subset\mathfrak{so}(n)$ that is normal.
	\item The partition $\lambda$ of the nilpotent orbit is given in exponential notation, i.e $[2^3]$ is the same as $[2,2,2]$. 
	\item Partitions $\lambda$ corresponding to non-special orbits and non-normal orbits are indicated explicitly. No quivers are provided for these orbits as we cannot obtain quivers whose Coulomb branch is the closure of the nilpotent orbit using the prescriptions in sections \ref{method} and \ref{math}.
	\item Trivial partitions $\lambda = [1^n]$ have been omitted from the analysis since the corresponding Coulomb branch is trivial.
    \item The \emph{very even} partition $\lambda=[2^2]$ corresponds to two different orbits ${\mathcal O}_{[2^2]}^I$ and ${\mathcal O}_{[2^2]}^{II}$ under the adjoint action of $SO(4)$. The corresponding Coulomb branch is isomorphic to the closure of only one of these orbits. This is also the case for the remaining very even partitions $\lambda=[2^4]$ and $\lambda=[4^2]$.
	\item The symbol ($\dagger$) denotes a quiver where the GNO magnetic lattice of two gauge groups is projected by a diagonal action of a $\mathbb Z_2$ group. This action is described in section \ref{sec:Z2actions}. 
	\item The Lusztig's Canonical Quotient\footnote{This is a finite discrete group associated to each nilpotent orbit. It was defined in \cite{L84} for special orbits and in \cite{S01} extended to all orbits. See also \cite{A02,CDT13}.} $\bar{A}(\mathcal{O}_\lambda)$ corresponding to each orbit is shown. The choice of $SO/O(N)$ gauge factors in the quiver seems to be beautifully encoded in this group. This is discussed at length in section \ref{sec:analysis}.
\end{itemize}

It is worthwhile to highlight that the resulting quivers with Coulomb branch $\mathcal{C}_1$ being the closure of the minimal orbit of $\mathfrak{so}(2n)$ (with partition $\lambda=[2^2,1^{2n-4}]$) and the resulting quivers with Coulomb branch $\mathcal{C}_2$ being the closure of the next to minimal orbit of $\mathfrak{so}(2n-1)$ (with partition $\lambda=[3,1^{2n-4}]$) differ only by a $\mathbb{Z}_2$ factor on the first node (this is the rightmost node in the tables). This is a compelling physical realization of the mathematical result by Brylinski and Kostant, that states that the two varieties are related by a $\mathbb Z_2$ quotient \cite{BK92}: $\mathcal C_2 = \mathcal C_1/\mathbb Z_2$.

\begin{table}[t]
			\centering
			\begin{tabular}{|c|c|c|}
				\hline
				Partition ($\lambda$)  &$T^\sigma(SO(3)^\vee)$ &$\bar{A}(\mathcal{O}_\lambda)$\\ \hline
				
				$[3]$

				 & \raisebox{-.5\height}{\begin{tikzpicture}
				\tikzstyle{gauge1} = [draw=none,minimum size=0.35cm,fill=blue,circle, draw];
				\tikzstyle{gauge2} = [draw=none,minimum size=0.35cm,fill=red,circle, draw];
				\tikzstyle{flavour1} = [draw=none,minimum size=0.35cm,fill=blue, regular polygon,regular polygon sides=4,draw];
				\tikzstyle{flavour2} = [draw=none,minimum size=0.35cm,fill=red, regular polygon,regular polygon sides=4,draw];
				\node (g1) [gauge2,label=below:{$2$}] {};
			
				\node (f1) [flavour1,above of=g1,label=above:{$2$}] {};
				
				\draw (g1)--(f1) 
				;
				\end{tikzpicture}                      }

				&1
				
				\\ \hline

			\end{tabular}
			
			\caption{$T^\sigma (SO(3)^\vee)$ quivers with Coulomb branch the closure of a nilpotent orbit $\bar{\mathcal O}_\lambda\subset\mathfrak{so}(3)$. The partition $\sigma=d_{BV}(\lambda)$ is the dual partition of $\lambda$ under the Barbasch-Vogan map. $\AO$ is the Lusztig's Canonical Quotient of each orbit.}
			\label{table1}
\end{table}

\begin{table}[t]
	\centering
	\begin{tabular}{|c|c|c|}
				\hline
				Partition ($\lambda$)  &$T^\sigma(SO(4)^\vee)$ &$\bar{A}(\mathcal{O}_\lambda)$\\ \hline
				
				$[3,1]$

				& \raisebox{-.5\height}{\begin{tikzpicture}
				\tikzstyle{gauge1} = [draw=none,minimum size=0.35cm,fill=blue,circle, draw];
				\tikzstyle{gauge2} = [draw=none,minimum size=0.35cm,fill=red,circle, draw];
				\tikzstyle{flavour1} = [draw=none,minimum size=0.35cm,fill=blue, regular polygon,regular polygon sides=4,draw];
				\tikzstyle{flavour2} = [draw=none,minimum size=0.35cm,fill=red, regular polygon,regular polygon sides=4,draw];
				\node (g1) [gauge2,label=below:{$2$}] {};
				\node (g2) [gauge1,right of=g1,label=below:{$2$}] {};
				\node (f6) [flavour2,above of=g2,label=above:{$4$}] {};

				\draw (g2)--(f6)
				(g1)--(g2) ;
				\end{tikzpicture}      }

				&1
				\\ \hline

				$[2^2]$

				  & \raisebox{-.5\height}{\begin{tikzpicture}
				\tikzstyle{gauge1} = [draw=none,minimum size=0.35cm,fill=blue,circle, draw];
				\tikzstyle{gauge2} = [draw=none,minimum size=0.35cm,fill=red,circle, draw];
				\tikzstyle{flavour1} = [draw=none,minimum size=0.35cm,fill=blue, regular polygon,regular polygon sides=4,draw];
				\tikzstyle{flavour2} = [draw=none,minimum size=0.35cm,fill=red, regular polygon,regular polygon sides=4,draw];
				\node (g1) [gauge2,label=below:{$2$}] {};
				\node (f6) [flavour1,above of=g1,label=above:{$2$}] {};

				\draw 
				(g1)--(f6) ;
				\end{tikzpicture}      }

				&1
				\\ \hline

			\end{tabular}
			
   			\caption{Quivers with $\mathcal C=\bar{\mathcal O}_\lambda\subset\mathfrak{so}(4)$. The dual partition is $\sigma=d_{BV}(\lambda)$.}
	
			\label{table2}
\end{table}

\begin{table}[t]
		\centering
		\begin{tabular}{|c|c|c|}
				\hline
				Partition ($\lambda$)  &$T^\sigma(SO(5)^\vee)$ &$\bar{A}(\mathcal{O}_\lambda)$\\ \hline
				
				$[5]$

				 & \raisebox{-.5\height}{\begin{tikzpicture}
				\tikzstyle{gauge1} = [draw=none,minimum size=0.35cm,fill=blue,circle, draw];
				\tikzstyle{gauge2} = [draw=none,minimum size=0.35cm,fill=red,circle, draw];
				\tikzstyle{flavour1} = [draw=none,minimum size=0.35cm,fill=blue, regular polygon,regular polygon sides=4,draw];
				\tikzstyle{flavour2} = [draw=none,minimum size=0.35cm,fill=red, regular polygon,regular polygon sides=4,draw];
				\node (g1) [gauge2,label=below:{$2$}] {};
				\node (g2) [gauge1,right of=g1,label=below:{$2$}] {};

				\node (g3) [gauge2,right of=g2,label=below:{$4$}] {};
				\node (f1) [flavour1,above of=g3,label=above:{$4$}] {};
				
				\draw (g3)--(f1) 
				(g1)--(g2) (g2)--(g3);
				\end{tikzpicture}      }

				&1
				
				\\ \hline

				$[3,1^2]$

				 & \raisebox{-.5\height}{\begin{tikzpicture}
				\tikzstyle{gauge1} = [draw=none,minimum size=0.35cm,fill=blue,circle, draw];
				\tikzstyle{gauge2} = [draw=none,minimum size=0.35cm,fill=red,circle, draw];
				\tikzstyle{flavour1} = [draw=none,minimum size=0.35cm,fill=blue, regular polygon,regular polygon sides=4,draw];
				\tikzstyle{flavour2} = [draw=none,minimum size=0.35cm,fill=red, regular polygon,regular polygon sides=4,draw];
				\node (g1) [gauge2,label=below:{$2$}] {};
				\node (g2) [gauge1,right of=g1,label=below:{$2$}] {};
				\node (g3) [gauge2,right of=g2,label=below:{$O(2)$}] {};		
				\node (f1) [flavour2,above of=g2,label=above:{$2$}] {};

				\draw (g2)--(f1) 
				(g1)--(g2) (g2)--(g3);
				\end{tikzpicture}      }

				&
				 $\mathbb{Z}_2$

				\\ \hline

				$[2^2,1]$&Non-special&1   \\ \hline

		\end{tabular}
        \bigskip
		\caption{Quivers with $\mathcal C=\bar{\mathcal O}_\lambda\subset\mathfrak{so}(5)$. The dual partition is $\sigma=d_{BV}(\lambda)$.}
        
        \label{table3}
		
\end{table}

\begin{table}[t]
		\centering
		\begin{tabular}{|c|c|c|}
				\hline
				Partition ($\lambda$)  &$T^\sigma(SO(6)^\vee)$ &$\bar{A}(\mathcal{O}_\lambda)$\\ \hline

				$[5,1]$

				 &\raisebox{-.5\height}{ \begin{tikzpicture}
				\tikzstyle{gauge1} = [draw=none,minimum size=0.35cm,fill=blue,circle, draw];
				\tikzstyle{gauge2} = [draw=none,minimum size=0.35cm,fill=red,circle, draw];
				\tikzstyle{flavour1} = [draw=none,minimum size=0.35cm,fill=blue, regular polygon,regular polygon sides=4,draw];
				\tikzstyle{flavour2} = [draw=none,minimum size=0.35cm,fill=red, regular polygon,regular polygon sides=4,draw];
				\node (g1) [gauge2,label=below:{$2$}] {};
				\node (g2) [gauge1,right of=g1,label=below:{$2$}] {};
				\node (g3) [gauge2,right of=g2,label=below:{$4$}] {};
				\node (g4) [gauge1,right of=g3,label=below:{$4$}] {};

				\node (f1) [flavour2,above of=g4,label=above:{$6$}] {};

				\draw (g4)--(f1)
				(g1)--(g2) (g2)--(g3) (g3)--(g4) ;
				\end{tikzpicture}            }

				&1
				
				\\ \hline

				$[3^2]$

				 & \raisebox{-.5\height}{\begin{tikzpicture}
				\tikzstyle{gauge1} = [draw=none,minimum size=0.35cm,fill=blue,circle, draw];
				\tikzstyle{gauge2} = [draw=none,minimum size=0.35cm,fill=red,circle, draw];
				\tikzstyle{flavour1} = [draw=none,minimum size=0.35cm,fill=blue, regular polygon,regular polygon sides=4,draw];
				\tikzstyle{flavour2} = [draw=none,minimum size=0.35cm,fill=red, regular polygon,regular polygon sides=4,draw];
				\node (g1) [gauge2,label=below:{$2$}] {};
				\node (g2) [gauge1,right of=g1,label=below:{$2$}] {};
				\node (g3) [gauge2,right of=g2,label=below:{$4$}] {};
				
				\node (g6) [gauge1,right of=g3,label=below:{$2$}] {};
				\node (f6) [flavour1,above of=g3,label=above:{$2$}] {};
				\node (f7) [flavour2,above of=g4,label=above:{$2$}] {};

				\draw (g3)--(f6) (g4)--(f7)
				(g1)--(g2) (g2)--(g3) (g3)--(g6)  ;
				\end{tikzpicture}    }

				&1

				\\ \hline

				$[3,1^3]$	
				
				 & \raisebox{-.5\height}{\begin{tikzpicture}
				\tikzstyle{gauge1} = [draw=none,minimum size=0.35cm,fill=blue,circle, draw];
				\tikzstyle{gauge2} = [draw=none,minimum size=0.35cm,fill=red,circle, draw];
				\tikzstyle{flavour1} = [draw=none,minimum size=0.35cm,fill=blue, regular polygon,regular polygon sides=4,draw];
				\tikzstyle{flavour2} = [draw=none,minimum size=0.35cm,fill=red, regular polygon,regular polygon sides=4,draw];
				\node (g1) [gauge2,label=below:{$2$}] {};
				\node (g2) [gauge1,right of=g1,label=below:{$2$}] {};
				\node (g3) [gauge2,right of=g2,label=below:{$3$}] {};
				\node (g4) [gauge1,right of=g3,label=below:{$2$}] {};
				
				\node (f1) [flavour2,above of=g2,label=above:{$1$}] {};
				\node (f2) [flavour2,above of=g4,label=above:{$3$}] {};
				
				\draw (g2)--(f1)  (g4)--(f2)  
				(g1)--(g2) (g2)--(g3) (g3)--(g4) ;
				\end{tikzpicture}      }

				&1

				\\ \hline

				$[2^2,1^2]$

				& \raisebox{-.5\height}{\begin{tikzpicture}
				\tikzstyle{gauge1} = [draw=none,minimum size=0.35cm,fill=blue,circle, draw];
				\tikzstyle{gauge2} = [draw=none,minimum size=0.35cm,fill=red,circle, draw];
				\tikzstyle{flavour1} = [draw=none,minimum size=0.35cm,fill=blue, regular polygon,regular polygon sides=4,draw];
				\tikzstyle{flavour2} = [draw=none,minimum size=0.35cm,fill=red, regular polygon,regular polygon sides=4,draw];
				\node (g1) [gauge2,label=below:{$2$}] {};
				\node (g2) [gauge1,right of=g1,label=below:{$2$}] {};
				\node (f2) [flavour2,above of=g2,label=above:{$2$}] {};
				\node (g3) [gauge2,right of=g2,label=below:{$2$}] {};

				\draw (g2)--(f2)
				(g1)--(g2) (g2)--(g3) ;
				\end{tikzpicture}    }

				&1
				
				\\ \hline

			\end{tabular}
			\caption{Quivers with $\mathcal C=\bar{\mathcal O}_\lambda\subset\mathfrak{so}(6)$. The dual partition is $\sigma=d_{BV}(\lambda)$.}
			\label{table4}
\end{table}

\begin{table}[t]
		\centering
		\begin{tabular}{|c|c|c|}
				\hline
				Partition ($\lambda$)  &$T^\sigma(SO(7)^\vee)$ &$\bar{A}(\mathcal{O}_\lambda)$\\ \hline
				
				$[7]$

				 & \raisebox{-.5\height}{\begin{tikzpicture}
				\tikzstyle{gauge1} = [draw=none,minimum size=0.35cm,fill=blue,circle, draw];
				\tikzstyle{gauge2} = [draw=none,minimum size=0.35cm,fill=red,circle, draw];
				\tikzstyle{flavour1} = [draw=none,minimum size=0.35cm,fill=blue, regular polygon,regular polygon sides=4,draw];
				\tikzstyle{flavour2} = [draw=none,minimum size=0.35cm,fill=red, regular polygon,regular polygon sides=4,draw];
				\node (g1) [gauge2,label=below:{$2$}] {};
				\node (g2) [gauge1,right of=g1,label=below:{$2$}] {};
				\node (g3) [gauge2,right of=g2,label=below:{$4$}] {};
				\node (g4) [gauge1,right of=g3,label=below:{$4$}] {};
				\node (g5) [gauge2,right of=g4,label=below:{$6$}] {};
				
				\node (f1) [flavour1,above of=g5,label=above:{$6$}] {};

				\draw (g5)--(f1) 
				(g1)--(g2) (g2)--(g3) (g3)--(g4) (g4)--(g5);

				\end{tikzpicture}     }

				&1
				\\ \hline

				$[5,1^2]$

				 & \raisebox{-.5\height}{\begin{tikzpicture}
				\tikzstyle{gauge1} = [draw=none,minimum size=0.35cm,fill=blue,circle, draw];
				\tikzstyle{gauge2} = [draw=none,minimum size=0.35cm,fill=red,circle, draw];
				\tikzstyle{flavour1} = [draw=none,minimum size=0.35cm,fill=blue, regular polygon,regular polygon sides=4,draw];
				\tikzstyle{flavour2} = [draw=none,minimum size=0.35cm,fill=red, regular polygon,regular polygon sides=4,draw];
				\node (g1) [gauge2,label=below:{$2$}] {};
				\node (g2) [gauge1,right of=g1,label=below:{$2$}] {};
				\node (g3) [gauge2,right of=g2,label=below:{$4$}] {};
				\node (g4) [gauge1,right of=g3,label=below:{$4$}] {};
				\node (g5) [gauge2,right of=g4,label=below:{$O(4)$}] {};
				
				\node (f1) [flavour2,above of=g4,label=above:{$2$}] {};
				\node (f2) [flavour1,above of=g5,label=above:{$2$}] {};		
				
				\draw (g4)--(f1) (g5)--(f2) 
				(g1)--(g2) (g2)--(g3) (g3)--(g4) (g4)--(g5);

				\draw (g4)--(f1)
				(g1)--(g2) (g2)--(g3) (g3)--(g4) ;
				\end{tikzpicture}     }

				& $\mathbb{Z}_2$
				
				\\ \hline
				
				$[3^2,1]$

				& \raisebox{-.5\height}{\begin{tikzpicture}
				\tikzstyle{gauge1} = [draw=none,minimum size=0.35cm,fill=blue,circle, draw];
				\tikzstyle{gauge2} = [draw=none,minimum size=0.35cm,fill=red,circle, draw];
				\tikzstyle{flavour1} = [draw=none,minimum size=0.35cm,fill=blue, regular polygon,regular polygon sides=4,draw];
				\tikzstyle{flavour2} = [draw=none,minimum size=0.35cm,fill=red, regular polygon,regular polygon sides=4,draw];
				\node (g1) [gauge2,label=below:{$2$}] {};
				\node (g2) [gauge1,right of=g1,label=below:{$2$}] {};
				\node (g3) [gauge2,right of=g2,label=below:{$4$}] {};
				\node (g4) [gauge1,right of=g3,label=below:{$4$}] {};
				\node (g5) [gauge2,right of=g4,label=below:{$3$}] {};
				
				\node (f1) [flavour2,above of=g4,label=above:{$3$}] {};

				\draw (g4)--(f1) 
				(g1)--(g2) (g2)--(g3) (g3)--(g4) (g4)--(g5);
				\end{tikzpicture}      }

				&1
				\\ \hline		
				$[3,2^2]$

				&Non-normal
			&1
				\\ \hline

				$[3,1^4]$

				 & \raisebox{-.5\height}{\begin{tikzpicture}
				\tikzstyle{gauge1} = [draw=none,minimum size=0.35cm,fill=blue,circle, draw];
				\tikzstyle{gauge2} = [draw=none,minimum size=0.35cm,fill=red,circle, draw];
				\tikzstyle{flavour1} = [draw=none,minimum size=0.35cm,fill=blue, regular polygon,regular polygon sides=4,draw];
				\tikzstyle{flavour2} = [draw=none,minimum size=0.35cm,fill=red, regular polygon,regular polygon sides=4,draw];
				\node (g1) [gauge2,label=below:{$2$}] {};
				\node (g2) [gauge1,right of=g1,label=below:{$2$}] {};
				\node (g3) [gauge2,right of=g2,label=below:{$3$}] {};
				\node (g4) [gauge1,right of=g3,label=below:{$2$}] {};
				\node (g5) [gauge2,right of=g4,label=below:{$O(2)$}] {};
				
				\node (f1) [flavour2,above of=g2,label=above:{$1$}] {};
				\node (f2) [flavour2,above of=g4,label=above:{$1$}] {};
				
				\draw (g2)--(f1) (g4)--(f2)
				(g1)--(g2) (g2)--(g3) (g3)--(g4) (g4)--(g5);
				\end{tikzpicture}     }

				& $\mathbb{Z}_2$
				
				\\ \hline

				$[2^2,1^3]$ &Non-special &1   \\ \hline			
			\end{tabular}
		
			\caption{$T^\sigma (SO(7)^\vee)$ quivers with Coulomb branch the closure of a nilpotent orbit $\bar{\mathcal O}_\lambda$ of $\mathfrak{so}(7)$. The partition $\sigma=d_{BV}(\lambda)$ is the dual partition of $\lambda$ under the Barbasch-Vogan map. $\AO$ is the Lusztig's Canonical Quotient of each orbit.}
			\label{table5}
\end{table}

\begin{table}[t]
	\centering
	\begin{tabular}{|c|c|c|}
				\hline
				Partition ($\lambda$)  &$T^\sigma(SO(8)^\vee)$ &$\bar{A}(\mathcal{O}_\lambda)$\\ \hline
				
				$[7,1]$

				  & \raisebox{-.5\height}{\begin{tikzpicture}
				\tikzstyle{gauge1} = [draw=none,minimum size=0.35cm,fill=blue,circle, draw];
				\tikzstyle{gauge2} = [draw=none,minimum size=0.35cm,fill=red,circle, draw];
				\tikzstyle{flavour1} = [draw=none,minimum size=0.35cm,fill=blue, regular polygon,regular polygon sides=4,draw];
				\tikzstyle{flavour2} = [draw=none,minimum size=0.35cm,fill=red, regular polygon,regular polygon sides=4,draw];
				\node (g1) [gauge2,label=below:{$2$}] {};
				\node (g2) [gauge1,right of=g1,label=below:{$2$}] {};
				\node (g3) [gauge2,right of=g2,label=below:{$4$}] {};
				\node (g4) [gauge1,right of=g3,label=below:{$4$}] {};
				\node (g5) [gauge2,right of=g4,label=below:{$6$}] {};
				\node (g6) [gauge1,right of=g5,label=below:{$6$}] {};
				
				\node (f1) [flavour2,above of=g6,label=above:{$8$}] {};
				\draw (g6)--(f1)   
				(g1)--(g2) (g2)--(g3) (g3)--(g4) (g4)--(g5) (g5)--(g6) ;
				\end{tikzpicture}  }

				&1

				\\ \hline	
				
				$[5,3]$

				 & \raisebox{-.5\height}{\begin{tikzpicture}
				\tikzstyle{gauge1} = [draw=none,minimum size=0.35cm,fill=blue,circle, draw];
				\tikzstyle{gauge2} = [draw=none,minimum size=0.35cm,fill=red,circle, draw];
				\tikzstyle{flavour1} = [draw=none,minimum size=0.35cm,fill=blue, regular polygon,regular polygon sides=4,draw];
				\tikzstyle{flavour2} = [draw=none,minimum size=0.35cm,fill=red, regular polygon,regular polygon sides=4,draw];
				\node (g1) [gauge2,label=below:{$2$}] {};
				\node (g2) [gauge1,right of=g1,label=below:{$2$}] {};
				\node (g3) [gauge2,right of=g2,label=below:{$4$}] {};
				\node (g4) [gauge1,right of=g3,label=below:{$4$}] {};
				\node (g5) [gauge2,right of=g4,label=below:{$6$}] {};
				\node (g6) [gauge1,right of=g5,label=below:{$4$}] {};
				
				\node (f1) [flavour1,above of=g5,label=above:{$2$}] {};
				\node (f2) [flavour2,above of=g6,label=above:{$4$}] {};		
				
				\draw (g5)--(f1)   (g6)--(f2)  
				(g1)--(g2) (g2)--(g3) (g3)--(g4) (g4)--(g5) (g5)--(g6) ;
				\end{tikzpicture}   }

				&1
				\\ 
				\hline
				$[5,1^3]$

				 & \raisebox{-.5\height}{\begin{tikzpicture}
				\tikzstyle{gauge1} = [draw=none,minimum size=0.35cm,fill=blue,circle, draw];
				\tikzstyle{gauge2} = [draw=none,minimum size=0.35cm,fill=red,circle, draw];
				\tikzstyle{flavour1} = [draw=none,minimum size=0.35cm,fill=blue, regular polygon,regular polygon sides=4,draw];
				\tikzstyle{flavour2} = [draw=none,minimum size=0.35cm,fill=red, regular polygon,regular polygon sides=4,draw];
				\node (g1) [gauge2,label=below:{$2$}] {};
				\node (g2) [gauge1,right of=g1,label=below:{$2$}] {};
				\node (g3) [gauge2,right of=g2,label=below:{$4$}] {};
				\node (g4) [gauge1,right of=g3,label=below:{$4$}] {};
				\node (g5) [gauge2,right of=g4,label=below:{$5$}] {};
				\node (g6) [gauge1,right of=g5,label=below:{$4$}] {};
				
				\node (f1) [flavour2,above of=g4,label=above:{$1$}] {};
				\node (f2) [flavour2,above of=g6,label=above:{$5$}] {};		
				
				\draw (g4)--(f1)   (g6)--(f2)  
				(g1)--(g2) (g2)--(g3) (g3)--(g4) (g4)--(g5) (g5)--(g6) ;
				\end{tikzpicture}     }

				&1

				\\ \hline

				$[4^2]$	
				
				 &\raisebox{-.5\height}{\begin{tikzpicture}
				\tikzstyle{gauge1} = [draw=none,minimum size=0.35cm,fill=blue,circle, draw];
				\tikzstyle{gauge2} = [draw=none,minimum size=0.35cm,fill=red,circle, draw];
				\tikzstyle{flavour1} = [draw=none,minimum size=0.35cm,fill=blue, regular polygon,regular polygon sides=4,draw];
				\tikzstyle{flavour2} = [draw=none,minimum size=0.35cm,fill=red, regular polygon,regular polygon sides=4,draw];
				\node (g1) [gauge2,label=below:{$2$}] {};
				\node (g2) [gauge1,right of=g1,label=below:{$2$}] {};
				\node (g3) [gauge2,right of=g2,label=below:{$4$}] {};
				\node (g4) [gauge1,right of=g3,label=below:{$4$}] {};
				\node (g5) [gauge2,right of=g4,label=below:{$6$}] {};
				\node (g6) [gauge1,right of=g5,label=below:{$2$}] {};
				
				\node (f1) [flavour1,above of=g5,label=above:{$4$}] {};
				
				\draw (g5)--(f1)  
				(g1)--(g2) (g2)--(g3) (g3)--(g4) (g4)--(g5) (g5)--(g6) ;
				\end{tikzpicture}   }   
				
				&1
				
				\\ \hline
				
				$[3^2,1^2]$
				
				 & \raisebox{-.5\height}{\begin{tikzpicture}
				\tikzstyle{gauge1} = [draw=none,minimum size=0.35cm,fill=blue,circle, draw];
				\tikzstyle{gauge2} = [draw=none,minimum size=0.35cm,fill=red,circle, draw];
				\tikzstyle{flavour1} = [draw=none,minimum size=0.35cm,fill=blue, regular polygon,regular polygon sides=4,draw];
				\tikzstyle{flavour2} = [draw=none,minimum size=0.35cm,fill=red, regular polygon,regular polygon sides=4,draw];
				\node (g1) [gauge2,label=below:{$2$}] {};
				\node (g2) [gauge1,right of=g1,label=below:{$2$}] {};
				\node (g3) [gauge2,right of=g2,label=below:{$4$}] {};
				\node (g4) [gauge1,right of=g3,label=below:{$4$}] {};
				\node (g5) [gauge2,right of=g4,label=below:{$O(4)$}] {};
				\node (g6) [gauge1,right of=g5,label=below:{$2$}] {};
				
				\node (f1) [flavour2,above of=g4,label=above:{$2$}] {};
				\node (f2) [flavour2,above of=g6,label=above:{$2$}] {};
				
				\draw (g4)--(f1)  (g6)--(f2)  
				(g1)--(g2) (g2)--(g3) (g3)--(g4) (g4)--(g5) (g5)--(g6) ;
				\end{tikzpicture}  }

				& $\mathbb{Z}_2$

				\\ \hline

			$[3,2^2,1]$ &Non-special&1   \\ \hline

				$[3,1^5]$

				 & \raisebox{-.5\height}{\begin{tikzpicture}
				\tikzstyle{gauge1} = [draw=none,minimum size=0.35cm,fill=blue,circle, draw];
				\tikzstyle{gauge2} = [draw=none,minimum size=0.35cm,fill=red,circle, draw];
				\tikzstyle{flavour1} = [draw=none,minimum size=0.35cm,fill=blue, regular polygon,regular polygon sides=4,draw];
				\tikzstyle{flavour2} = [draw=none,minimum size=0.35cm,fill=red, regular polygon,regular polygon sides=4,draw];
				\node (g1) [gauge2,label=below:{$2$}] {};
				\node (g2) [gauge1,right of=g1,label=below:{$2$}] {};
				\node (g3) [gauge2,right of=g2,label=below:{$3$}] {};
				\node (g4) [gauge1,right of=g3,label=below:{$2$}] {};
				\node (g5) [gauge2,right of=g4,label=below:{$3$}] {};
				\node (g6) [gauge1,right of=g5,label=below:{$2$}] {};
				
				\node (f1) [flavour2,above of=g2,label=above:{$1$}] {};
				\node (f2) [flavour2,above of=g6,label=above:{$3$}] {};
				
				\draw (g2)--(f1)  (g6)--(f2)  
				(g1)--(g2) (g2)--(g3) (g3)--(g4) (g4)--(g5) (g5)--(g6) ;
				\end{tikzpicture}     }

				&1

				\\ \hline

				$[2^4]$
				
				 &\raisebox{-.5\height}{\begin{tikzpicture}
				\tikzstyle{gauge1} = [draw=none,minimum size=0.35cm,fill=blue,circle, draw];
				\tikzstyle{gauge2} = [draw=none,minimum size=0.35cm,fill=red,circle, draw];
				\tikzstyle{flavour1} = [draw=none,minimum size=0.35cm,fill=blue, regular polygon,regular polygon sides=4,draw];
				\tikzstyle{flavour2} = [draw=none,minimum size=0.35cm,fill=red, regular polygon,regular polygon sides=4,draw];
				\node (g1) [gauge2,label=below:{$2$}] {};
				\node (g2) [gauge1,right of=g1,label=below:{$2$}] {};
				\node (f3) [flavour1,above of=g3,label=above:{$2$}] {};
				\node (g3) [gauge2,right of=g2,label=below:{$4$}] {};
				\node (g6) [gauge1,right of=g3,label=below:{$2$}] {};
				\node (g7) [gauge2,right of=g6,label=below:{$2$}] {};
				
				\draw (g3)--(f3)
				(g1)--(g2) (g2)--(g3) (g3)--(g6)  (g6)--(g7);
				\end{tikzpicture}      }        
				&1

				\\ \hline

				$[2^2,1^4]$

				 & \raisebox{-.5\height}{\begin{tikzpicture}
				\tikzstyle{gauge1} = [draw=none,minimum size=0.35cm,fill=blue,circle, draw];
				\tikzstyle{gauge2} = [draw=none,minimum size=0.35cm,fill=red,circle, draw];
				\tikzstyle{flavour1} = [draw=none,minimum size=0.35cm,fill=blue, regular polygon,regular polygon sides=4,draw];
				\tikzstyle{flavour2} = [draw=none,minimum size=0.35cm,fill=red, regular polygon,regular polygon sides=4,draw];
				\node (g1) [gauge2,label=below:{$2$}] {};
				\node (g2) [gauge1,right of=g1,label=below:{$2$}] {};
				\node (f2) [flavour2,above of=g2,label=above:{$1$}] {};
				\node (g3) [gauge2,right of=g2,label=below:{$3$}] {};
				\node (g6) [gauge1,right of=g3,label=below:{$2$}] {};
				\node (f6) [flavour2,above of=g6,label=above:{$1$}] {};
				\node (g7) [gauge2,right of=g6,label=below:{$2$}] {};
				
				\draw (g2)--(f2) (g6)--(f6)
				(g1)--(g2) (g2)--(g3) (g3)--(g6)  (g6)--(g7);
				\end{tikzpicture}      }

				&1
				
				\\ \hline

			\end{tabular}

			\caption{Quivers with $\mathcal C=\bar{\mathcal O}_\lambda\subset\mathfrak{so}(8)$. The dual partition is $\sigma=d_{BV}(\lambda)$.}
			\label{table6}
\end{table}

\begin{table}[t]
		\centering
		\begin{tabular}{|c|c|c|}
				\hline
				Partition ($\lambda$)  &$T^\sigma(SO(9)^\vee)$ &$\bar{A}(\mathcal{O}_\lambda)$\\ \hline

				$[9]$

				  & \raisebox{-.5\height}{\begin{tikzpicture}
				\tikzstyle{gauge1} = [draw=none,minimum size=0.35cm,fill=blue,circle, draw];
				\tikzstyle{gauge2} = [draw=none,minimum size=0.35cm,fill=red,circle, draw];
				\tikzstyle{flavour1} = [draw=none,minimum size=0.35cm,fill=blue, regular polygon,regular polygon sides=4,draw];
				\tikzstyle{flavour2} = [draw=none,minimum size=0.35cm,fill=red, regular polygon,regular polygon sides=4,draw];
				\node (g1) [gauge2,label=below:{$2$}] {};
				\node (g2) [gauge1,right of=g1,label=below:{$2$}] {};
				\node (g3) [gauge2,right of=g2,label=below:{$4$}] {};
				\node (g4) [gauge1,right of=g3,label=below:{$4$}] {};
				\node (g5) [gauge2,right of=g4,label=below:{$6$}] {};
				\node (g6) [gauge1,right of=g5,label=below:{$6$}] {};
				\node (g7) [gauge2,right of=g6,label=below:{$8$}] {};
				
				\node (f1) [flavour1,above of=g7,label=above:{$8$}] {};

				\draw (g7)--(f1) 
				(g1)--(g2) (g2)--(g3) (g3)--(g4) (g4)--(g5) (g5)--(g6)  (g6)--(g7);
				\end{tikzpicture}    }

				&1

				\\ \hline

				$[7,1^2]$

				 & \raisebox{-.5\height}{\begin{tikzpicture}
				\tikzstyle{gauge1} = [draw=none,minimum size=0.35cm,fill=blue,circle, draw];
				\tikzstyle{gauge2} = [draw=none,minimum size=0.35cm,fill=red,circle, draw];
				\tikzstyle{flavour1} = [draw=none,minimum size=0.35cm,fill=blue, regular polygon,regular polygon sides=4,draw];
				\tikzstyle{flavour2} = [draw=none,minimum size=0.35cm,fill=red, regular polygon,regular polygon sides=4,draw];
				\node (g1) [gauge2,label=below:{$2$}] {};
				\node (g2) [gauge1,right of=g1,label=below:{$2$}] {};
				\node (g3) [gauge2,right of=g2,label=below:{$4$}] {};
				\node (g4) [gauge1,right of=g3,label=below:{$4$}] {};
				\node (g5) [gauge2,right of=g4,label=below:{$6$}] {};
				\node (g6) [gauge1,right of=g5,label=below:{$6$}] {};
				\node (g7) [gauge2,right of=g6,label=below:{$O(6)$}] {};
				
				\node (f1) [flavour1,above of=g7,label=above:{$4$}] {};
				\node (f2) [flavour2,above of=g6,label=above:{$2$}] {};		
				
				\draw (g7)--(f1) (g6)--(f2)
				(g1)--(g2) (g2)--(g3) (g3)--(g4) (g4)--(g5) (g5)--(g6)  (g6)--(g7);
				\end{tikzpicture}    }

				& $\mathbb{Z}_2$
				
				\\ \hline

				$[5,3,1]$

				 & \raisebox{-.5\height}{\begin{tikzpicture} 
				\tikzstyle{gauge1} = [draw=none,minimum size=0.35cm,fill=blue,circle, draw];
				\tikzstyle{gauge2} = [draw=none,minimum size=0.35cm,fill=red,circle, draw];
				\tikzstyle{flavour1} = [draw=none,minimum size=0.35cm,fill=blue, regular polygon,regular polygon sides=4,draw];
				\tikzstyle{flavour2} = [draw=none,minimum size=0.35cm,fill=red, regular polygon,regular polygon sides=4,draw];
				\node (g1) [gauge2,label=below:{$2$}] {};
				\node (g2) [gauge1,right of=g1,label=below:{$2$}] {};
				\node (g3) [gauge2,right of=g2,label=below:{$4$}] {};
				\node (g4) [gauge1,right of=g3,label=below:{$4$}] {};
				\node (g5) [gauge2,right of=g4,label=below:{$6$}] {};
				\node (g6) [gauge1,right of=g5,label=below:{$6$}] {};
				\node (g7) [gauge2,right of=g6,label=below:{$O(4)$}] {};
				
				\node (f1) [flavour2,above of=g6,label=above:{$4$}] {};

				\draw (g6)--(f1) 
				(g1)--(g2) (g2)--(g3) (g3)--(g4) (g4)--(g5) (g5)--(g6)  (g6)--(g7);
				\end{tikzpicture}    }

				& $\mathbb{Z}_2$

				\\ \hline

				$[5,2^2]$

				 & Non-normal

				&1

				\\ \hline		 
					$[4^2,1]$ &Non-special&1\\ \hline
				
				$[5,1^4]$

				 & \raisebox{-.5\height}{\begin{tikzpicture}
				\tikzstyle{gauge1} = [draw=none,minimum size=0.35cm,fill=blue,circle, draw];
				\tikzstyle{gauge2} = [draw=none,minimum size=0.35cm,fill=red,circle, draw];
				\tikzstyle{flavour1} = [draw=none,minimum size=0.35cm,fill=blue, regular polygon,regular polygon sides=4,draw];
				\tikzstyle{flavour2} = [draw=none,minimum size=0.35cm,fill=red, regular polygon,regular polygon sides=4,draw];
				\node (g1) [gauge2,label=below:{$2$}] {};
				\node (g2) [gauge1,right of=g1,label=below:{$2$}] {};
				\node (g3) [gauge2,right of=g2,label=below:{$4$}] {};
				\node (g4) [gauge1,right of=g3,label=below:{$4$}] {};
				\node (g5) [gauge2,right of=g4,label=below:{$5$}] {};
				\node (g6) [gauge1,right of=g5,label=below:{$4$}] {};
				\node (g7) [gauge2,right of=g6,label=below:{$O(4)$}] {};
				
				\node (f1) [flavour2,above of=g4,label=above:{$1$}] {};
				\node (f2) [flavour2,above of=g6,label=above:{$1$}] {};		
				\node (f3) [flavour1,above of=g7,label=above:{$2$}] {};			
				\draw (g4)--(f1) (g6)--(f2) (g7)--(f3)
				(g1)--(g2) (g2)--(g3) (g3)--(g4) (g4)--(g5) (g5)--(g6)  (g6)--(g7);
				\end{tikzpicture}     }

				& $\mathbb{Z}_2$

				\\ \hline

			\end{tabular}
			
			\caption{$T^\sigma (SO(9)^\vee)$ quivers with Coulomb branch the closure of a nilpotent orbit $\bar{\mathcal O}_\lambda$ of $\mathfrak{so}(9)$. The partition $\sigma=d_{BV}(\lambda)$ is the dual partition of $\lambda$ under the Barbasch-Vogan map. $\AO$ is the Lusztig's Canonical Quotient of each orbit.}
			\label{table7}
\end{table}

		\begin{table}[t]
			\centering
			
		\begin{tabular}{|c|c|c|}
				\hline
				Partition ($\lambda$)  &$T^\sigma(SO(9)^\vee)$ &$\bar{A}(\mathcal{O}_\lambda)$\\ \hline

			$[3^3]$

				& \raisebox{-.5\height}{\begin{tikzpicture}
				\tikzstyle{gauge1} = [draw=none,minimum size=0.35cm,fill=blue,circle, draw];
				\tikzstyle{gauge2} = [draw=none,minimum size=0.35cm,fill=red,circle, draw];
				\tikzstyle{flavour1} = [draw=none,minimum size=0.35cm,fill=blue, regular polygon,regular polygon sides=4,draw];
				\tikzstyle{flavour2} = [draw=none,minimum size=0.35cm,fill=red, regular polygon,regular polygon sides=4,draw];
				\node (g1) [gauge2,label=below:{$2$}] {};
				\node (g2) [gauge1,right of=g1,label=below:{$2$}] {};
				\node (g3) [gauge2,right of=g2,label=below:{$4$}] {};
				\node (g4) [gauge1,right of=g3,label=below:{$4$}] {};
				\node (g5) [gauge2,right of=g4,label=below:{$6$}] {};
				\node (g6) [gauge1,right of=g5,label=below:{$4$}] {};
				\node (g7) [gauge2,right of=g6,label=below:{$3$}] {};
				
				\node (f1) [flavour1,above of=g5,label=above:{$2$}] {};
				\node (f2) [flavour2,above of=g6,label=above:{$1$}] {};		
				
				\draw (g5)--(f1) (g6)--(f2)
				(g1)--(g2) (g2)--(g3) (g3)--(g4) (g4)--(g5) (g5)--(g6)  (g6)--(g7);
				\end{tikzpicture}    }

				&1
				
				\\ \hline

				$[3^2,1^3]$

				 & \raisebox{-.5\height}{\begin{tikzpicture}
				\tikzstyle{gauge1} = [draw=none,minimum size=0.35cm,fill=blue,circle, draw];
				\tikzstyle{gauge2} = [draw=none,minimum size=0.35cm,fill=red,circle, draw];
				\tikzstyle{flavour1} = [draw=none,minimum size=0.35cm,fill=blue, regular polygon,regular polygon sides=4,draw];
				\tikzstyle{flavour2} = [draw=none,minimum size=0.35cm,fill=red, regular polygon,regular polygon sides=4,draw];
				\node (g1) [gauge2,label=below:{$2$}] {};
				\node (g2) [gauge1,right of=g1,label=below:{$2$}] {};
				\node (g3) [gauge2,right of=g2,label=below:{$4$}] {};
				\node (g4) [gauge1,right of=g3,label=below:{$4$}] {};
				\node (g5) [gauge2,right of=g4,label=below:{$5$}] {};
				\node (g6) [gauge1,right of=g5,label=below:{$4$}] {};
				\node (g7) [gauge2,right of=g6,label=below:{$3$}] {};
				
				\node (f1) [flavour2,above of=g4,label=above:{$1$}] {};
				\node (f2) [flavour2,above of=g6,label=above:{$2$}] {};		
				
				\draw (g4)--(f1) (g6)--(f2)
				(g1)--(g2) (g2)--(g3) (g3)--(g4) (g4)--(g5) (g5)--(g6)  (g6)--(g7);
				\end{tikzpicture}       }

				&1
				
				\\ \hline

					$[3,2^2,1^2]^\dagger $

				& \raisebox{-.5\height}{\begin{tikzpicture}
				\tikzstyle{gauge1} = [draw=none,minimum size=0.35cm,fill=blue,circle, draw];
				\tikzstyle{gauge2} = [draw=none,minimum size=0.35cm,fill=red,circle, draw];
				\tikzstyle{flavour1} = [draw=none,minimum size=0.35cm,fill=blue, regular polygon,regular polygon sides=4,draw];
				\tikzstyle{flavour2} = [draw=none,minimum size=0.35cm,fill=red, regular polygon,regular polygon sides=4,draw];
				\node (g1) [gauge2,label=below:{$2$}] {};
				\node (g2) [gauge1,right of=g1,label=below:{$2$}] {};
				\node (g3) [gauge2,right of=g2,label=below:{$4$}] {};
				
				\node (g4) [gauge1,right of=g3,label=below:{$4$}] {};
				\node (f6) [flavour2,above of=g4,label=above:{$2$}] {};
				\node (g5) [gauge2,right of=g4,label=below:{$[4$}] {};
				\node (g6) [gauge1,right of=g5,label=below:{$2$}] {};
				\node (g7) [gauge2,right of=g6,label=below:{$2]$}] {};		
				\draw (g4)--(f6)
				(g1)--(g2) (g2)--(g3) (g3)--(g4) (g4)--(g5) (g5)--(g6)  (g6)--(g7);
				\end{tikzpicture}   }

				&$\mathbb{Z}_2$
				\\ \hline
					$[2^4,1]$ &Non-special &1  \\ \hline

				$[3,1^6]$

				 & \raisebox{-.5\height}{\begin{tikzpicture}
				\tikzstyle{gauge1} = [draw=none,minimum size=0.35cm,fill=blue,circle, draw];
				\tikzstyle{gauge2} = [draw=none,minimum size=0.35cm,fill=red,circle, draw];
				\tikzstyle{flavour1} = [draw=none,minimum size=0.35cm,fill=blue, regular polygon,regular polygon sides=4,draw];
				\tikzstyle{flavour2} = [draw=none,minimum size=0.35cm,fill=red, regular polygon,regular polygon sides=4,draw];
				\node (g1) [gauge2,label=below:{$2$}] {};
				\node (g2) [gauge1,right of=g1,label=below:{$2$}] {};
				\node (f2) [flavour2,above of=g2,label=above:{$1$}] {};
				\node (g3) [gauge2,right of=g2,label=below:{$3$}] {};
				\node (g4) [gauge1,right of=g3,label=below:{$2$}] {};
				\node (g5) [gauge2,right of=g4,label=below:{$3$}] {};
				\node (g6) [gauge1,right of=g5,label=below:{$2$}] {};
				\node (f6) [flavour2,above of=g6,label=above:{$1$}] {};
				\node (g7) [gauge2,right of=g6,label=below:{$O(2)$}] {};
				
				\draw (g2)--(f2) (g6)--(f6)
				(g1)--(g2) (g2)--(g3) (g3)--(g4)  (g4)--(g5)  (g5)--(g6)  (g6)--(g7);
				\end{tikzpicture} }

				& $\mathbb{Z}_2$
				\\ \hline	
				
				$[2^2,1^5]$ &Non-special &1  \\ \hline
			\end{tabular}
			
			\caption{Quivers with $\mathcal C=\bar{\mathcal O}_\lambda\subset\mathfrak{so}(9)$. The dual partition is $\sigma=d_{BV}(\lambda)$. The brackets in row $[3,2^2,1^2]^\dagger$ correspond to a choice of gauge nodes discussed in section \ref{sec:Z2actions}.}
			\label{table8}
\end{table}

\begin{table}[t]
			\centering
			\begin{tabular}{|c|c|c|}
				\hline
				Partition ($\lambda$)  &$T^\sigma(SO(10)^\vee)$ &$\bar{A}(\mathcal{O}_\lambda)$\\ \hline
				
				$[9,1]$

				& \raisebox{-.5\height}{\begin{tikzpicture}
				\tikzstyle{gauge1} = [draw=none,minimum size=0.35cm,fill=blue,circle, draw];
				\tikzstyle{gauge2} = [draw=none,minimum size=0.35cm,fill=red,circle, draw];
				\tikzstyle{flavour1} = [draw=none,minimum size=0.35cm,fill=blue, regular polygon,regular polygon sides=4,draw];
				\tikzstyle{flavour2} = [draw=none,minimum size=0.35cm,fill=red, regular polygon,regular polygon sides=4,draw];
				\node (g1) [gauge2,label=below:{$2$}] {};
				\node (g2) [gauge1,right of=g1,label=below:{$2$}] {};
				\node (g3) [gauge2,right of=g2,label=below:{$4$}] {};
				\node (g4) [gauge1,right of=g3,label=below:{$4$}] {};
				\node (g5) [gauge2,right of=g4,label=below:{$6$}] {};
				\node (g6) [gauge1,right of=g5,label=below:{$6$}] {};
				\node (g7) [gauge2,right of=g6,label=below:{$8$}] {};
				\node (g8) [gauge1,right of=g7,label=below:{$8$}] {};
				
				\node (f2) [flavour2,above of=g8,label=above:{$10$}] {};
				
				\draw  (g8)--(f2) 
				(g1)--(g2) (g2)--(g3) (g3)--(g4) (g4)--(g5) (g5)--(g6) (g6)--(g7) (g7)--(g8) ;
				\end{tikzpicture}}               
				&1
				
				\\ \hline
				
				$[7,3]$			
				
				 & \raisebox{-.5\height}{\begin{tikzpicture}
				\tikzstyle{gauge1} = [draw=none,minimum size=0.35cm,fill=blue,circle, draw];
				\tikzstyle{gauge2} = [draw=none,minimum size=0.35cm,fill=red,circle, draw];
				\tikzstyle{flavour1} = [draw=none,minimum size=0.35cm,fill=blue, regular polygon,regular polygon sides=4,draw];
				\tikzstyle{flavour2} = [draw=none,minimum size=0.35cm,fill=red, regular polygon,regular polygon sides=4,draw];
				\node (g1) [gauge2,label=below:{$2$}] {};
				\node (g2) [gauge1,right of=g1,label=below:{$2$}] {};
				\node (g3) [gauge2,right of=g2,label=below:{$4$}] {};
				\node (g4) [gauge1,right of=g3,label=below:{$4$}] {};
				\node (g5) [gauge2,right of=g4,label=below:{$6$}] {};
				\node (g6) [gauge1,right of=g5,label=below:{$6$}] {};
				\node (g7) [gauge2,right of=g6,label=below:{$8$}] {};
				\node (g8) [gauge1,right of=g7,label=below:{$6$}] {};
				\node (f1) [flavour1,above of=g7,label=above:{$2$}] {};
				\node (f2) [flavour2,above of=g8,label=above:{$6$}] {};
				
				\draw (g7)--(f1)  (g8)--(f2) 
				(g1)--(g2) (g2)--(g3) (g3)--(g4) (g4)--(g5) (g5)--(g6) (g6)--(g7) (g7)--(g8) ;
				\end{tikzpicture}    }

				&1

				\\ \hline

				$[5,5]$			
				
				 & \raisebox{-.5\height}{\begin{tikzpicture}
				\tikzstyle{gauge1} = [draw=none,minimum size=0.35cm,fill=blue,circle, draw];
				\tikzstyle{gauge2} = [draw=none,minimum size=0.35cm,fill=red,circle, draw];
				\tikzstyle{flavour1} = [draw=none,minimum size=0.35cm,fill=blue, regular polygon,regular polygon sides=4,draw];
				\tikzstyle{flavour2} = [draw=none,minimum size=0.35cm,fill=red, regular polygon,regular polygon sides=4,draw];
				\node (g1) [gauge2,label=below:{$2$}] {};
				\node (g2) [gauge1,right of=g1,label=below:{$2$}] {};
				\node (g3) [gauge2,right of=g2,label=below:{$4$}] {};
				\node (g4) [gauge1,right of=g3,label=below:{$4$}] {};
				\node (g5) [gauge2,right of=g4,label=below:{$6$}] {};
				\node (g6) [gauge1,right of=g5,label=below:{$6$}] {};
				\node (g7) [gauge2,right of=g6,label=below:{$8$}] {};
				\node (g8) [gauge1,right of=g7,label=below:{$4$}] {};
				\node (f1) [flavour1,above of=g7,label=above:{$4$}] {};
				\node (f2) [flavour2,above of=g8,label=above:{$2$}] {};
				
				\draw (g7)--(f1)  (g8)--(f2) 
				(g1)--(g2) (g2)--(g3) (g3)--(g4) (g4)--(g5) (g5)--(g6) (g6)--(g7) (g7)--(g8) ;
				\end{tikzpicture}      }

				&1
				\\ \hline

				$[7,1^3]$		
				
				  & \raisebox{-.5\height}{\begin{tikzpicture}
				\tikzstyle{gauge1} = [draw=none,minimum size=0.35cm,fill=blue,circle, draw];
				\tikzstyle{gauge2} = [draw=none,minimum size=0.35cm,fill=red,circle, draw];
				\tikzstyle{flavour1} = [draw=none,minimum size=0.35cm,fill=blue, regular polygon,regular polygon sides=4,draw];
				\tikzstyle{flavour2} = [draw=none,minimum size=0.35cm,fill=red, regular polygon,regular polygon sides=4,draw];
				\node (g1) [gauge2,label=below:{$2$}] {};
				\node (g2) [gauge1,right of=g1,label=below:{$2$}] {};
				\node (g3) [gauge2,right of=g2,label=below:{$4$}] {};
				\node (g4) [gauge1,right of=g3,label=below:{$4$}] {};
				\node (g5) [gauge2,right of=g4,label=below:{$6$}] {};
				\node (g6) [gauge1,right of=g5,label=below:{$6$}] {};
				\node (g7) [gauge2,right of=g6,label=below:{$7$}] {};
				\node (g8) [gauge1,right of=g7,label=below:{$6$}] {};
				\node (f1) [flavour2,above of=g6,label=above:{$1$}] {};
				\node (f2) [flavour2,above of=g8,label=above:{$7$}] {};
				
				\draw (g6)--(f1)  (g8)--(f2) 
				(g1)--(g2) (g2)--(g3) (g3)--(g4) (g4)--(g5) (g5)--(g6) (g6)--(g7) (g7)--(g8) ;
				\end{tikzpicture}      }

				&1
				
				\\ \hline

				$[5,3,1^2]$
				
				 & \raisebox{-.5\height}{\begin{tikzpicture}
				\tikzstyle{gauge1} = [draw=none,minimum size=0.35cm,fill=blue,circle, draw];
				\tikzstyle{gauge2} = [draw=none,minimum size=0.35cm,fill=red,circle, draw];
				\tikzstyle{flavour1} = [draw=none,minimum size=0.35cm,fill=blue, regular polygon,regular polygon sides=4,draw];
				\tikzstyle{flavour2} = [draw=none,minimum size=0.35cm,fill=red, regular polygon,regular polygon sides=4,draw];
				\node (g1) [gauge2,label=below:{$2$}] {};
				\node (g2) [gauge1,right of=g1,label=below:{$2$}] {};
				\node (g3) [gauge2,right of=g2,label=below:{$4$}] {};
				\node (g4) [gauge1,right of=g3,label=below:{$4$}] {};
				\node (g5) [gauge2,right of=g4,label=below:{$6$}] {};
				\node (g6) [gauge1,right of=g5,label=below:{$6$}] {};
				\node (g7) [gauge2,right of=g6,label=below:{$O(6)$}] {};
				\node (g8) [gauge1,right of=g7,label=below:{$4$}] {};
				\node (f1) [flavour2,above of=g6,label=above:{$2$}] {};
				\node (f2) [flavour2,above of=g8,label=above:{$4$}] {};
				
				\draw (g6)--(f1)  (g8)--(f2) 
				(g1)--(g2) (g2)--(g3) (g3)--(g4) (g4)--(g5) (g5)--(g6) (g6)--(g7) (g7)--(g8) ;
				\end{tikzpicture}          }

				& $\mathbb{Z}_2$
				\\ \hline		
				
				$[4^2,1^2]$
				
				& \raisebox{-.5\height}{\begin{tikzpicture}
				\tikzstyle{gauge1} = [draw=none,minimum size=0.35cm,fill=blue,circle, draw];
				\tikzstyle{gauge2} = [draw=none,minimum size=0.35cm,fill=red,circle, draw];
				\tikzstyle{flavour1} = [draw=none,minimum size=0.35cm,fill=blue, regular polygon,regular polygon sides=4,draw];
				\tikzstyle{flavour2} = [draw=none,minimum size=0.35cm,fill=red, regular polygon,regular polygon sides=4,draw];
				\node (g1) [gauge2,label=below:{$2$}] {};
				\node (g2) [gauge1,right of=g1,label=below:{$2$}] {};
				\node (g3) [gauge2,right of=g2,label=below:{$4$}] {};
				\node (g4) [gauge1,right of=g3,label=below:{$4$}] {};
				\node (g5) [gauge2,right of=g4,label=below:{$6$}] {};
				\node (g6) [gauge1,right of=g5,label=below:{$6$}] {};
				\node (g7) [gauge2,right of=g6,label=below:{$6$}] {};
				\node (g8) [gauge1,right of=g7,label=below:{$2$}] {};
				\node (f1) [flavour2,above of=g6,label=above:{$2$}] {};
				\node (f2) [flavour1,above of=g7,label=above:{$2$}] {};
				
				\draw (g6)--(f1)  (g7)--(f2) 
				(g1)--(g2) (g2)--(g3) (g3)--(g4) (g4)--(g5) (g5)--(g6) (g6)--(g7) (g7)--(g8) ;
				\end{tikzpicture}    }

				&1

				\\ \hline

				$[5,2^2,1]$ &Non-special&   \\ \hline

				$[3^3,1]$		
				 & \raisebox{-.5\height}{\begin{tikzpicture}
				\tikzstyle{gauge1} = [draw=none,minimum size=0.35cm,fill=blue,circle, draw];
				\tikzstyle{gauge2} = [draw=none,minimum size=0.35cm,fill=red,circle, draw];
				\tikzstyle{flavour1} = [draw=none,minimum size=0.35cm,fill=blue, regular polygon,regular polygon sides=4,draw];
				\tikzstyle{flavour2} = [draw=none,minimum size=0.35cm,fill=red, regular polygon,regular polygon sides=4,draw];
				\node (g1) [gauge2,label=below:{$2$}] {};
				\node (g2) [gauge1,right of=g1,label=below:{$2$}] {};
				\node (g3) [gauge2,right of=g2,label=below:{$4$}] {};
				\node (g4) [gauge1,right of=g3,label=below:{$4$}] {};
				\node (g5) [gauge2,right of=g4,label=below:{$6$}] {};
				\node (g6) [gauge1,right of=g5,label=below:{$6$}] {};
				\node (g7) [gauge2,right of=g6,label=below:{$5$}] {};
				\node (g8) [gauge1,right of=g7,label=below:{$2$}] {};
				\node (f1) [flavour2,above of=g6,label=above:{$3$}] {};
				\node (f2) [flavour2,above of=g8,label=above:{$1$}] {};
				
				\draw (g6)--(f1)  (g8)--(f2) 
				(g1)--(g2) (g2)--(g3) (g3)--(g4) (g4)--(g5) (g5)--(g6) (g6)--(g7) (g7)--(g8) ;
				\end{tikzpicture}  }

				&1
				\\ \hline

			\end{tabular}
			
			\caption{Quivers with $\mathcal C=\bar{\mathcal O}_\lambda\subset\mathfrak{so}(10)$. The dual partition is $\sigma=d_{BV}(\lambda)$.} 
			\label{table9}
\end{table}

\begin{table}[t]
		\centering

		\begin{tabular}{|c|c|c|}
				\hline
				Partition ($\lambda$)  &$T^\sigma(SO(10)^\vee)$ &$\bar{A}(\mathcal{O}_\lambda)$\\ \hline

				$[5,1^5]$
				
				 & \raisebox{-.5\height}{\begin{tikzpicture}
				\tikzstyle{gauge1} = [draw=none,minimum size=0.35cm,fill=blue,circle, draw];
				\tikzstyle{gauge2} = [draw=none,minimum size=0.35cm,fill=red,circle, draw];
				\tikzstyle{flavour1} = [draw=none,minimum size=0.35cm,fill=blue, regular polygon,regular polygon sides=4,draw];
				\tikzstyle{flavour2} = [draw=none,minimum size=0.35cm,fill=red, regular polygon,regular polygon sides=4,draw];
				\node (g1) [gauge2,label=below:{$2$}] {};
				\node (g2) [gauge1,right of=g1,label=below:{$2$}] {};
				\node (g3) [gauge2,right of=g2,label=below:{$4$}] {};
				\node (g4) [gauge1,right of=g3,label=below:{$4$}] {};
				\node (g5) [gauge2,right of=g4,label=below:{$5$}] {};
				\node (g6) [gauge1,right of=g5,label=below:{$4$}] {};
				\node (g7) [gauge2,right of=g6,label=below:{$5$}] {};
				\node (g8) [gauge1,right of=g7,label=below:{$4$}] {};
				\node (f1) [flavour2,above of=g4,label=above:{$1$}] {};
				\node (f2) [flavour2,above of=g8,label=above:{$5$}] {};
				
				\draw (g4)--(f1)  (g8)--(f2) 
				(g1)--(g2) (g2)--(g3) (g3)--(g4) (g4)--(g5) (g5)--(g6) (g6)--(g7) (g7)--(g8) ;
				\end{tikzpicture} }

				&1
				\\ \hline

				$[3^2,2^2]$	
				 & Non-normal
				&1
				\\ \hline

				$[3^2,1^4]$	
				  & \raisebox{-.5\height}{\begin{tikzpicture}
				\tikzstyle{gauge1} = [draw=none,minimum size=0.35cm,fill=blue,circle, draw];
				\tikzstyle{gauge2} = [draw=none,minimum size=0.35cm,fill=red,circle, draw];
				\tikzstyle{flavour1} = [draw=none,minimum size=0.35cm,fill=blue, regular polygon,regular polygon sides=4,draw];
				\tikzstyle{flavour2} = [draw=none,minimum size=0.35cm,fill=red, regular polygon,regular polygon sides=4,draw];
				\node (g1) [gauge2,label=below:{$2$}] {};
				\node (g2) [gauge1,right of=g1,label=below:{$2$}] {};
				\node (g3) [gauge2,right of=g2,label=below:{$4$}] {};
				\node (g4) [gauge1,right of=g3,label=below:{$4$}] {};
				\node (g5) [gauge2,right of=g4,label=below:{$5$}] {};
				\node (g6) [gauge1,right of=g5,label=below:{$4$}] {};
				\node (g7) [gauge2,right of=g6,label=below:{$O(4)$}] {};
				\node (g8) [gauge1,right of=g7,label=below:{$2$}] {};
				\node (f1) [flavour2,above of=g4,label=above:{$1$}] {};
				\node (f2) [flavour2,above of=g6,label=above:{$1$}] {};
				\node (f3) [flavour2,above of=g8,label=above:{$2$}] {};
				\draw (g4)--(f1)  (g6)--(f2)  (g8)--(f3)
				(g1)--(g2) (g2)--(g3) (g3)--(g4) (g4)--(g5) (g5)--(g6) (g6)--(g7) (g7)--(g8) ;
				\end{tikzpicture}       }

				& $\mathbb{Z}_2$ 
				
			\\	\hline

				$[3,2^2,1^3]$ &Non-special &1 \\ \hline

				$[2^4,1^2]$	
				
				 & \raisebox{-.5\height}{\begin{tikzpicture}
				\tikzstyle{gauge1} = [draw=none,minimum size=0.35cm,fill=blue,circle, draw];
				\tikzstyle{gauge2} = [draw=none,minimum size=0.35cm,fill=red,circle, draw];
				\tikzstyle{flavour1} = [draw=none,minimum size=0.35cm,fill=blue, regular polygon,regular polygon sides=4,draw];
				\tikzstyle{flavour2} = [draw=none,minimum size=0.35cm,fill=red, regular polygon,regular polygon sides=4,draw];
				\node (g1) [gauge2,label=below:{$2$}] {};
				\node (g2) [gauge1,right of=g1,label=below:{$2$}] {};
				\node (g3) [gauge2,right of=g2,label=below:{$4$}] {};
				\node (g4) [gauge1,right of=g3,label=below:{$4$}] {};
				\node (g5) [gauge2,right of=g4,label=below:{$4$}] {};
				\node (g6) [gauge1,right of=g5,label=below:{$2$}] {};
				\node (f6) [flavour2,above of=g4,label=above:{$2$}] {};
				\node (g7) [gauge2,right of=g6,label=below:{$2$}] {};
				
				\draw (g4)--(f6)
				(g1)--(g2) (g2)--(g3) (g3)--(g4) (g4)--(g5) (g5)--(g6) (g6)--(g7);
				\end{tikzpicture}   }

				&1

				\\ \hline

				$[3,1^7]$	 
				
				& \raisebox{-.5\height}{\begin{tikzpicture}
				\tikzstyle{gauge1} = [draw=none,minimum size=0.35cm,fill=blue,circle, draw];
				\tikzstyle{gauge2} = [draw=none,minimum size=0.35cm,fill=red,circle, draw];
				\tikzstyle{flavour1} = [draw=none,minimum size=0.35cm,fill=blue, regular polygon,regular polygon sides=4,draw];
				\tikzstyle{flavour2} = [draw=none,minimum size=0.35cm,fill=red, regular polygon,regular polygon sides=4,draw];
				\node (g1) [gauge2,label=below:{$2$}] {};
				\node (g2) [gauge1,right of=g1,label=below:{$2$}] {};
				\node (f2) [flavour2,above of=g2,label=above:{$1$}] {};
				\node (g3) [gauge2,right of=g2,label=below:{$3$}] {};
				\node (g4) [gauge1,right of=g3,label=below:{$2$}] {};
				\node (g5) [gauge2,right of=g4,label=below:{$3$}] {};
				\node (g6) [gauge1,right of=g5,label=below:{$2$}] {};
				\node (g7) [gauge2,right of=g6,label=below:{$3$}] {};
				\node (g8) [gauge1,right of=g7,label=below:{$2$}] {};
				\node (f8) [flavour2,above of=g8,label=above:{$3$}] {};
				\draw (g2)--(f2)
				(g1)--(g2) (g2)--(g3) (g3)--(g4) (g4)--(g5) (g5)--(g6) (g6)--(g7) (g7)--(g8)  (g8)--(f8);
				\end{tikzpicture}   }              
				
				&1

				\\ \hline

				$[2^2,1^6]$			
				 &\raisebox{-.5\height}{ \begin{tikzpicture}
				\tikzstyle{gauge1} = [draw=none,minimum size=0.35cm,fill=blue,circle, draw];
				\tikzstyle{gauge2} = [draw=none,minimum size=0.35cm,fill=red,circle, draw];
				\tikzstyle{flavour1} = [draw=none,minimum size=0.35cm,fill=blue, regular polygon,regular polygon sides=4,draw];
				\tikzstyle{flavour2} = [draw=none,minimum size=0.35cm,fill=red, regular polygon,regular polygon sides=4,draw];
				\node (g1) [gauge2,label=below:{$2$}] {};
				\node (g2) [gauge1,right of=g1,label=below:{$2$}] {};
				\node (f2) [flavour2,above of=g2,label=above:{$1$}] {};
				\node (g3) [gauge2,right of=g2,label=below:{$3$}] {};
				\node (g4) [gauge1,right of=g3,label=below:{$2$}] {};
				\node (g5) [gauge2,right of=g4,label=below:{$3$}] {};
				\node (g6) [gauge1,right of=g5,label=below:{$2$}] {};
				\node (f6) [flavour2,above of=g6,label=above:{$1$}] {};
				\node (g7) [gauge2,right of=g6,label=below:{$2$}] {};
				
				\draw (g2)--(f2) (g6)--(f6)
				(g1)--(g2) (g2)--(g3) (g3)--(g4) (g4)--(g5) (g5)--(g6) (g6)--(g7);
				\end{tikzpicture}  }

				&1
				\\ \hline

			\end{tabular}
			\caption{Quivers with $\mathcal C=\bar{\mathcal O}_\lambda\subset\mathfrak{so}(10)$. The dual partition is $\sigma=d_{BV}(\lambda)$.} 
			\label{table10}
\end{table}

\clearpage

\subsection{Tables: Hilbert series and plethystic logarithms}
 
The Hilbert series of the Coulomb branch of the set of $T^\sigma(G^\vee)$ quivers presented in the previous section are computed using the \emph{monopole formula} (see section \ref{math}). 

The computation of the Hilbert series $H(t)$ is done perturbatively. $H(t)$ is expanded in powers of $t$ around $t=0$:
\begin{align}
	H(t)=\sum_{i=0}^{\infty}a_it^i
\end{align}
and the coefficients $a_i$ are computed starting with $a_0=1$. The highest order $a_k$ that can be computed is restricted by the computing power, obtaining an expression:
\begin{align}
	H(t)=\sum_{i=0}^{k}a_it^i+O(t^{k+1})
\end{align}

One can then use properties of the nilpotent orbit such as its complex dimension (which in this case equals twice the rank of the gauge group of the corresponding $T^\sigma (G^\vee)$ quiver).  Closures of nilpotent orbits that are normal have a Hilbert series with rational form:

\begin{align}
	H(t)=\frac{P(t)}{(1-t)^p}
\end{align}
where $p$ is the complex dimension and $P(t)$ is a palindromic polynomial of degree $p/2$ with $P(1)$ non zero. This allows to reconstruct the rational form of $H(t)$ after computing a finite set of coefficients $a_i$.

For nilpotent orbits of $\mathfrak{so}(n)$ with $n<10$ the Hilbert series of closures of nilpotent orbits can be compared with the results in \cite{HK16}. For nilpotent orbits of $\mathfrak{so}(10)$ we turn our attention to the plethystic logarithm defined in section \ref{math}, which provides the degrees of generators and the degrees of relations between them. This information can be compared with the constrains (such as \emph{rank} or \emph{degree of nilpotency}) on the nilpotent orbit when viewed as a set of matrices, i.e. linear operators acting on the fundamental representation of the Lie algebra\footnote{We have also compared the Hilbert series obtained for $\mathfrak{so}(10)$ with Hilbert series for closures of nilpotent orbits computed via localization techniques by Rudolph Kalveks (these results are described in an upcoming publication \cite{HK17}).}.

The following tables include the Hilbert series and the plethystic logarithms of the quivers displayed in the previous section. The plethystic logarithm is given up to an order so it at least includes the relations that show nilpotency.

\begin{table}[t]

	\centering
	\begin{tabular}{|c|c|c|}
		\hline
		Partition ($\lambda$)   &Hilbert series  &Plethystic logarithm\\ \hline

		$[3]$	               & 

$
		\dfrac{(1 + t)}{(1 - t)^2}
$

		 & \parbox[t]{3.5cm}{$3t-t^2$}

			\\ \hline

	\end{tabular}
	\caption{Hilbert series and plethystic logarithm for the Coulomb branch of quivers in table \ref{table1}.}
	\label{tab:HS3}
\end{table}

\begin{table}[t]
	\centering
	\begin{tabular}{|c|c|c|}
		\hline
		Partition ($\lambda$)   &Hilbert series  &Plethystic logarithm\\ \hline
	
			$[3,1]$	&

$	\dfrac{(1 - t^2)^2}{(1 - t)^6}  $

		 & \parbox[t]{3.5cm}{$6 t-2 t^2$}

		\\ \hline

		 $[2^2]$	&

$	\dfrac{1 + t}{(1 - t)^2}  $

		 & \parbox[t]{3.5cm}{$3 t- t^2$}

		\\ \hline

		\end{tabular}
		\caption{Hilbert series and plethystic logarithm for the Coulomb branch of quivers in table \ref{table2}.}
		\label{tab:HS4}
\end{table}

\begin{table}[t]

	\centering
	\begin{tabular}{|c|c|c|}
		\hline
		Partition ($\lambda$)    &Hilbert series  &Plethystic logarithm\\ \hline

		$[5]$	               & 

$
		\dfrac{(1 + t)^2 (1 + t^2)}{(1 - t)^8}
$

		 & \parbox[t]{3.5cm}{$10 t-t^2-t^4$}

			\\ \hline

	\raisebox{-.5\height}{	$[3,1^2]$ }          & 
		
	\raisebox{-.5\height}{
		  $ \dfrac{(1 + t) (1 + 3 t + t^2)}{(1 - t)^6} $}

	 &

			\parbox[t]{3.5cm}{$10 t-6 t^2+5 t^3-10 t^4+24 t^5+O(t^{6})$}

			\\ \hline

		$[2^2,1]$ &\multicolumn{2}{|c|}{Non-special}\\ \hline

	\end{tabular}
	
		\caption{Hilbert series and plethystic logarithm for the Coulomb branch of quivers in table \ref{table3}.}
		\label{tab:HS5}
\end{table}

\begin{table}[t]
	\centering
	\begin{tabular}{|c|c|c|}
		\hline
		Partition ($\lambda$)   &Hilbert series  &Plethystic logarithm\\ \hline

		\raisebox{-.5\height}{$[5,1]$}	&

$\dfrac{(1 - t^2) (1 - t^3) (1 - t^4)}{(1-t)^{15}}$

		 & \parbox[t]{3.5cm}{$15 t-t^2-t^3-t^4$}

		  \\ \hline

	\raisebox{-0.8\height}{	$[3^2]$	}&
		
		\raisebox{-.5\height}{
$\dfrac{1 + 4 t + 10 t^2 + 4 t^3 + t^4}{(1 - t)^{10} (1 + t)^{-1}}$}

		& \parbox[t]{3.5cm}{$15 t-t^2-16 t^3+30 t^4-16 t^5+O(t^{6})$}

		\\ \hline

		\raisebox{-.8\height}{$[3,1^3]$}	&
		
	\raisebox{-.5\height}{
$	\dfrac{1 + 5 t + t^2}{(1 - t)^8 (1 + t)^{-2}}$}

		 & \parbox[t]{3.5cm}{$15 t-16 t^2+35 t^3-126 t^4+504 t^5+O(t^6)$}

		 \\ \hline

	\raisebox{-.8\height}{	$[2^2,1^2]$	}&
		\raisebox{-.5\height}{
	$	\dfrac{1 + 8 t + t^2}{(1 - t)^6 (1 + t)^{-1}} $   
	}

		&\parbox[t]{3.5cm}{$15 t-36 t^2+160 t^3-945 t^4+6048 t^5+O(t^6)$}

		\\ \hline

	\end{tabular}

		\caption{Hilbert series and plethystic logarithm for the Coulomb branch of quivers in table \ref{table4}.}
		\label{tab:HS6}
\end{table}

\begin{table}[t]
    
	\centering
	\begin{tabular}{|c|c|c|}

		\hline
		Partition ($\lambda$)   &Hilbert series  &Plethystic logarithm\\ \hline
	
			$[7]$        &

		   $ \dfrac{(1 - t^2) (1 - t^4) (1 - t^6)}{(1 - t)^{21}}$

		 & \parbox[t]{3.5cm}{$21 t-t^2-t^4-t^6$}

			\\ \hline

	\raisebox{-.5\height}{	$[5,1^2]$	}&

	\raisebox{-.5\height}{$	\dfrac{1 + 3 t + 6 t^2 + 3 t^3 + t^4}{(1 - t)^{16} (1 + t)^{-2} (1 + t^2)^{-1}}    $}

	 &
	 \parbox[t]{3.5cm}{$21 t-t^2-7 t^3+6 t^4+O(t^6)$}

	 \\ \hline
	
		\raisebox{-.5\height}{$[3^2,1]$}&
		
	\raisebox{-.5\height}{$	 \dfrac{1 + 6 t + 21 t^2 + 28 t^3 + 21 t^4 + 6 t^5 + t^6}{(1 - t)^{14} (1 + t)^{-1}}   $}

		& \parbox[t]{3.5cm}{$21 t-t^2-28 t^3+63 t^4-36 t^5+O(t^6)$}

		\\ \hline

		\raisebox{-.5\height}{$[3,2^2]$  }     	&\multicolumn{2}{|c|}{Non-normal}

	 \\ \hline

		\raisebox{-.5\height}{$[3,1^4]$}&
		
\raisebox{-.5\height}{
	
$\dfrac{1+10t+20t^2+10t^3+t^4}{(1 - t)^{10}(1+t)^{-1}}$}

		&		
			\parbox[t]{3.5cm}{$21t-36 t^2+140 t^3-784 t^4+4788 t^5+O(t^6)$}

	\\ \hline

		$[2^2,1^3]$ &\multicolumn{2}{|c|}{Non-special}
		 \\ \hline

	\end{tabular}
	
		\caption{Hilbert series and plethystic logarithm for the Coulomb branch of quivers in table \ref{table5}.}
		\label{tab:HS7}
\end{table}

\begin{table}[t]
	\begin{adjustbox}{center}
	\begin{tabular}{|c|c|c|}
		\hline
		Partition ($\lambda$)   &Hilbert series  &Plethystic logarithm\\ \hline

		$[7,1]$	&

		$	\dfrac{(1 - t^2) (1 - t^4)^2 (1 - t^6)}{(1-t)^{28}}  $  

	 & \parbox[t]{3cm}{$28 t-t^2-2t^4-t^6$}

	 \\ \hline	

		\raisebox{-.5\height}{$[5,3]$	}&

	\raisebox{-.5\height}{$	\dfrac{1 + 3 t + 8 t^2 + 16 t^3 + 28 t^4 + 16 t^5 + 8 t^6 + 3 t^7 + t^8}{(1 - t)^{22} (1 + t)^{-3}}   $}

	 & \parbox[t]{3cm}{$28 t-t^2-2 t^4-28 t^5+O(t^6)$}

	 \\ \hline

		\raisebox{-.5\height}{$[5,1^3]$}	&

	\raisebox{-.5\height}{	$ \dfrac{1 + 6 t + 21 t^2 + 28 t^3 + 21 t^4 + 6 t^5 + t^6}{(1 - t)^{20} (1 + t)^{-2} (1 + t^2)^{-1}}  $ }

		 &\parbox[t]{3cm}{$28 t-t^2-28 t^3+62 t^4-36 t^5+O(t^6)$}

		 \\ \hline

		\raisebox{-.5\height}{$[4,4]$	}&

	\raisebox{-.5\height}{	$ \dfrac{1 + 6 t + 21 t^2 + 28 t^3 + 21 t^4 + 6 t^5 + t^6}{(1 - t)^{20} (1 + t)^{-2} (1 + t^2)^{-1}}  $ }

		 &\parbox[t]{3cm}{$28 t-t^2-28 t^3+62 t^4-36 t^5+O(t^6)$}

		 \\ \hline

		$[3^2,1^2]$	&
		\begin{tabular}{l}

$	\dfrac{1 + 9 t + 45 t^2 + 109 t^3 + 152 t^4 + 109 t^5 + 45 t^6 + 9 t^7 + t^8}{(1 - t)^{18} (1 + t)^{-1}} $

		\\

		\end{tabular}
		  &
		\begin{tabular}{l}
		 \parbox[t]{3cm}{ $28 t-t^2-56 t^3+161 t^4-107 t^5+O(t^6)$}\\

\end{tabular}

		\\ \hline

		$[3,2^2,1]$ &\multicolumn{2}{|c|}{Non-special}
		 \\ \hline

	\raisebox{-1\height}{$[3,1^5]$}	& 
		
		\raisebox{-.8\height}{	
$	\dfrac{1 + 14 t + 36 t^2 + 14 t^3 + t^4}{(1 - t)^{12} (1 + t)^{-2}}   $}

	 & \parbox[t]{3.5cm}{$28 t-71 t^2+420 t^3-3360 t^4+29148 t^5+O(t^6)$}

	 \\ \hline

	\raisebox{-1\height}{	$[2^4]$}	&
		
		\raisebox{-.8\height}{	
$	\dfrac{1 + 14 t + 36 t^2 + 14 t^3 + t^4}{(1 - t)^{12} (1 + t)^{-2}}   $}

	 & \parbox[t]{3.5cm}{$28 t-71 t^2+420 t^3-3360 t^4+29148 t^5+O(t^6)$}

	 \\ \hline

\raisebox{-1\height}{		$[2^2,1^4]$}	&
	\raisebox{-.8\height}{
	$	\dfrac{1+17t+48t^2+17t^3+t^4}{(1-t)^{10}(1+t)^{-1}  } $ }

	&\parbox[t]{3cm}{$28 t-106 t^2+833 t^3-8400 t^4+91392 t^5+O(t^6)$}

	\\ \hline

	\end{tabular}
	\end{adjustbox}
	
		\caption{Hilbert series and plethystic logarithm for the Coulomb branch of quivers in table \ref{table6}.}
		\label{tab:HS8}
\end{table}

\begin{center}
\begin{table}[t]

\begin{adjustbox}{center}
	\begin{tabular}{|c|c|c|}
		\hline
		Partition ($\lambda$)   &Hilbert series  &Plethystic logarithm\\ \hline

	$[9]$	& 
 
  $ 	\dfrac{(1 - t^2) (1 - t^4) (1 - t^6) (1 - t^8)}{(1-t)^{36}} $

	&

		\parbox[t]{3.5cm}{ $36 t-t^2-t^4-t^6$}

		  \\ \hline

	$[7,1^2]$	&

	$ \dfrac{
	 \splitfrac{1 + 5 t + 15 t^2 + 35 t^3 + 60 t^4 + 85 t^5 + 104 t^6 + 110 t^7 }
	 {+104 t^8 + 85 t^9 +60t^{10}+35t^{11}+15t^{12}+5t^{13}+t^{14}}
	 }{(1 - t)^{30} (1+t)^{-1}} $

	 & 
	 
		  \begin{tabular}{l}
		  \parbox[t]{3.5cm}{$36 t-t^2-10 t^4+9 t^5-t^6-36t^8+O(t^9)$}\\

		  \end{tabular}

		  \\ \hline

	$[5,3,1]$	&

	$ \dfrac{
	 \splitfrac{1 + 6 t + 22 t^2 + 62 t^3 + 138 t^4 + 227 t^5 + 264 t^6 }
	 { + 227 t^7 + 
			138 t^8 + 62 t^9 +22t^{10}+6t^{11}+t^{12}}
	 }{(1 - t)^{28} (1 + t)^{-2}}$

	&

		  \begin{tabular}{l}
		  \parbox[t]{3.5cm}{$36 t-t^2-10 t^4-27 t^5+36 t^6+O(t^7)$}\\

		  \end{tabular}

		  \\ \hline

	$[5,2^2]$	&\multicolumn{2}{|c|}{Non-normal}

 \\ \hline

		$[5,1^4]$	&

	$ \dfrac{
	 \splitfrac{1 + 10 t + 55 t^2 + 136 t^3 + 190 t^4}
	 { + 136 t^5 + 55 t^6 + 10 t^7 + t^8}
	 }{(1 - t)^{24} (1 + t)^{-2} (1 + t^2)^{-1}}  	 $

			 &\begin{tabular}{l}
			 \parbox[t]{3.5cm}{$36 t-t^2-84 t^3+314 t^4-396 t^5+O(t^6)$}\\

			 \end{tabular}

			 \\ \hline

		\raisebox{-.5\height}{$[3^3]$}	&

		\raisebox{-.5\height}{
	$ \dfrac{
	 \splitfrac{1 + 10 t + 56 t^2 + 194 t^3 + 438 t^4 + 578 t^5}
	 {  + 438 t^6 + 194 t^7 + 56 t^8 + 10 t^9 + t^{10}}
	 }{(1 - t)^{24} (1 + t)^{-2}}  	 $   }

		& \parbox[t]{3.5cm}{$36 t-t^2-36 t^3+27 t^4+92 t^5+O(t^6)$}

		\\ \hline

		\raisebox{-.5\height}{$[3^2,1^3]$}	&
			
		\raisebox{-.5\height}{
	$ \dfrac{
	 \splitfrac{1 + 13 t + 91 t^2 + 335 t^3 + 737 t^4 + 946 t^5}
	 { + 737 t^6 + 
			335 t^7 + 91t^8 + 13 t^9 + t^{10}}
	 }{(1 - t)^{22} (1 + t)^{-1}}  $	    }

			& \parbox[t]{3.5cm}{$36 t- t^2-120 t^3+477t^4-523t^5+O(t^6)$}

			\\ \hline
		$[3,2^2,1]$ &\multicolumn{2}{|c|}{Non-special}
		\\ \hline

		\raisebox{-1\height}{$[3,2^2,1^2]^\dagger $	}&
		
		\raisebox{-.8\height}{
					
		  $	\dfrac{\splitfrac{1+14t+106t^2+454t^3+788t^4}{+454t^5+106t^6+14t^7+t^8}}{(1-t)^{20}(1+t)^{-2}}$}

		 & \parbox[t]{3.5cm}{$36t-t^2-120t^3-18t^4+4284t^5+O(t^6)$}

		 \\ \hline

		$[3,1^6]$	&
		
		\begin{tabular}{l}
\raisebox{-.8\height}{
		$\dfrac{1 + 21 t + 105 t^2 + 175 t^3 + 105 t^4 + 21 t^5 + t^6}{(1 - t)^{14} (1 + t)^{-1}}$ }

		\end{tabular}		
	
			 &\begin{tabular}{l}
			 	\parbox[t]{3.5cm}{$36 t-127 t^2+1050 t^3-11340 t^4+132552 t^5+O(t^6)$}\\

			 \end{tabular}

			 \\ \hline	
			
		$[2^2,1^5]$ &\multicolumn{2}{|c|}{Non-special}
		 \\ \hline
	\end{tabular}
	\end{adjustbox}
	
		\caption{Hilbert series and plethystic logarithm for the Coulomb branch of quivers in table \ref{table7} and table \ref{table8}.}
		\label{tab:HS9}
\end{table}
\end{center}

\begin{table}[t]

\begin{adjustbox}{center}
	\begin{tabular}{|P{1.38cm}|P{10.8cm}|P{3cm}|}
		\hline
		Partition ($\lambda$)   &Hilbert series  &Plethystic logarithm\\ \hline

\raisebox{-1\height}{	$[9,1]$	}&

		\raisebox{-.5\height}{
	$\dfrac{(1-t^2)(1-t^4)(1-t^5)(1-t^6)(1-t^8)}{(1-t)^{45}}   $}

 & \parbox[t]{3cm}{$45t-t^2-t^4-t^5-t^6-t^8$}
	
	\\ \hline

		\raisebox{-.5\height}{$[7,3]$}	&
		
		\raisebox{-.5\height}{
	 $\dfrac{
	 \splitfrac{1+4t+12t^2+28t^3+57t^4+ 103t^5+171t^6+219t^7+250t^8}
	 {
	+219t^9+171t^{10}+103t^{11}+57t^{12}+28t^{13}+12t^{14}+4t^{15}+t^{16}}
	 }{(1-t)^{38}(1+t)^{-3}} $ 	 }

		& \parbox[t]{3cm}{$45 t-t^2-t^4-t^5-t^6-45 t^7+O(t^8)$} 
		\\ \hline

		\raisebox{-1.5\height}{$[5,5]$}	&
		
\raisebox{-1\height}{	
$	 \dfrac{
	 \splitfrac{1+7t+28t^2+84t^3+210t^4+ 416t^5+647t^6+742t^7}
	 {
		+647t^8+416t^9+210t^{10}+84t^{11}+28t^{12}+7t^{13}+t^{14}}
	 }{(1-t)^{36}(1+t)^{-2}(1+t^2)^{-1}}  $	  }

		& \parbox[t]{3cm}{$45t-t^2-t^4-46 t^5+O(t^6)$}

		\\ \hline

		\raisebox{-1.5\height}{$[7,1^3]$	}&
			
	\raisebox{-1\height}{
	 $\dfrac{
	 \splitfrac{1 + 5 t + 18 t^2 + 50 t^3 + 75 t^4 + 130 t^5 + 131 t^6 + 170 t^7}
	 {
			+131 t^8 + 130 t^9 + 75 t^{10} + 50 t^{11} + 18 t^{12} + 5 t^{13} + t^{14}}
	 }{(1 - t)^{36} (1 + t)^{-4}}  	$   }

			& 	\parbox[t]{3cm}{$45t-t^2-46t^4+99t^5- 56 t^6+O(t^8)$} 
			
			\\ \hline

		\raisebox{-1\height}{$[5,3,1^2]$	}
		&	
			\raisebox{-.6\height}{
	 $\dfrac{
	 \splitfrac{1+8t+38t^2+136t^3+359t^4+742t^5+1162t^6+1378t^7}
	 {+1162t^8+742t^9+359t^{10}+136t^{11}+38t^{12}+8t^{13}+t^{14}}
	 }{(1-t)^{34}(1+t)^{-3}}  $	   }

			&

		\parbox[t]{3cm}{$45t-t^2-46 t^4+54 t^5+O(t^6)$}

	\\ \hline		
	
		\raisebox{-1\height}{$[4^2,1^2]$}	&
		
		\raisebox{-.7\height}{
$	 \dfrac{
	 \splitfrac{1+11t+67t^2+297t^3+968t^2+2353t^5+4004t^6+4838t^7}
	 {+4004t^8+2353t^9+968t^{10}+297t^{11}+67t^{12}+11t^{13}+t^{14}}
	 }{(1-t)^{32}(1+t)^{-2}}  	$    }

		& \parbox[t]{3cm}{$45t-t^2-100 t^4+153 t^5+O(t^6)$}

		\\ \hline

		$[5,2^2,1]$&\multicolumn{2}{|c|}{Non-special} 
		\\ \hline

	\raisebox{-1\height}{	$[3^3,1]$	}& 
		
	\raisebox{-.5\height}{	
	$ \dfrac{
	 \splitfrac{1+14t+105t^2+515t^3+1750t^4+4151t^5+6924t^6+8200t^7}
	 {+6924t^8+4151t^9+1750t^{10}+515t^{11}+105t^{12}+14t^{13}+t^{14}}
	 }{(1-t)^{30}(1+t)^{-1}} $ 	 }

		& \parbox[t]{3cm}{$45t-t^2-45 t^3+308 t^5+O(t^6)$} 
		\\ \hline

	\end{tabular}
	\end{adjustbox}
	
	\caption{Hilbert series and plethystic logarithm for the Coulomb branch of quivers in table \ref{table9}.}
	\label{tab:HS101}
\end{table}

\begin{table}[t]

\begin{adjustbox}{center}
	\begin{tabular}{|P{1.38cm}|P{10.8cm}|P{3cm}|}
		\hline
		Partition ($\lambda$)   &Hilbert series  &Plethystic logarithm\\ \hline

	\raisebox{-1\height}{$[5,1^5]$	}&

\raisebox{-.8\height}{
$	 \dfrac{
	 \splitfrac{1+15t+121t^2+485t^3+1185t^4+1847t^5+2130t^6}
	 {+1847t^7+1185t^8+485t^9+121t^{10}+15t^{11}+t^{12}}
	 }{(1-t)^{28}(1+t)^{-2}}  	 $   }

	& \parbox[t]{3cm}{$45t-t^2-210 t^3+1154 t^4-2376 t^5+O(t^6)$}
	
	\\ \hline

	\raisebox{-.5\height}{$[3^2,2^2]$}	&\multicolumn{2}{|c|}{Non-normal}

		\\ \hline

		\raisebox{-.5\height}{$[3^2,1^4]$}	&

\raisebox{-.5\height}{

	$ \dfrac{
	 \splitfrac{1+18t+171t^2+885t^3+2805t^4+5522t^5+6936t^6}
	 {
+5522t^7+2805t^8+885t^9+171t^{10}+18t^{11}+t^{12}}
	 }{(1-t)^{26}(1+t)^{-1}}  $	    }
	    
		&
	
			\parbox[t]{3cm}{ $45 t-t^2-255 t^3+1410 t^4-2587 t^5+O(t^6)$}

		\\

			\hline	

		$[3,2^2,1^3]$&\multicolumn{2}{|c|}{Non-special}
		 \\ \hline

		\raisebox{-1\height}{$[2^4,1^2]$}	&
		
	\raisebox{-.8\height}{	
					
		  $	\dfrac{1+23t+223t^2+925t^3+1532t^4+925t^5+223t^6+23t^7+t^8}{(1-t)^{20}(1+t)^{-2}}$}

		 & \parbox[t]{3cm}{$45 t-55 t^2-156 t^3+3420 t^4-33471 t^5+O(t^6)$}

		 \\ \hline

		\raisebox{-1\height}{$[3,1^7]$}	&
		
	\raisebox{-.8\height}{		
		  $	\dfrac{1+27t+169t^2+321t^3+169t^4+27t^5+t^6}{(1-t)^{16}(1+t)^{-2}}$
}

		 & \parbox[t]{3cm}{$45t-211 t^2+2310 t^3-32340 t^4+489720 t^5+O(t^6)$}

		 \\ \hline

	\raisebox{-1\height}{$[2^2,1^6]$}	& 
\raisebox{-.8\height}{
		  $	\dfrac{1+30t+201t^2+394t^3+201t^4+30t^5+t^6}{(1-t)^{14}(1+t)^{-1}}$}

	 &\parbox[t]{3cm}{$45t-265t^2+3354t^3-53295t^4+914430t^5+O(t^6)$} 
	 \\ \hline

	\end{tabular}
	\end{adjustbox}
	
	\caption{Hilbert series and plethystic logarithm for the Coulomb branch of quivers in table \ref{table10}.}
	\label{tab:HS102}
\end{table}

\clearpage

\subsection{First case with $\AO=\mathbb Z_2^2$}

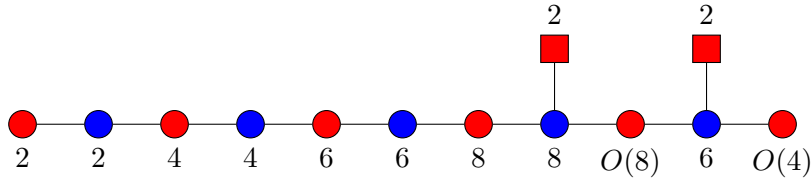
\begin{figure}[t]
	\centering
	\begin{tikzpicture}
		\node (g1) [gauge2,label=below:{$2$}] {};
		\node (g2) [gauge1,right of=g1,label=below:{$2$}] {};
		\node (g3) [gauge2,right of=g2,label=below:{$4$}] {};
		\node (g4) [gauge1,right of=g3,label=below:{$4$}] {};
		\node (g5) [gauge2,right of=g4,label=below:{$6$}] {};
		\node (g6) [gauge1,right of=g5,label=below:{$6$}] {};
		\node (g7) [gauge2,right of=g6,label=below:{$8$}] {};
		\node (g8) [gauge1,right of=g7,label=below:{$8$}] {};
		\node (g9) [gauge2,right of=g8,label=below:{$O(8)$}] {};
		\node (g10) [gauge1,right of=g9,label=below:{$6$}] {};
		\node (g11) [gauge2,right of=g10,label=below:{$O(4)$}] {};
		\node (f1) [flavour2,above of=g8,label=above:{$2$}] {};
		\node (f2) [flavour2,above of=g10,label=above:{$2$}] {};
		\draw (g10)--(f2) (g8)--(f1)
			(g1)--(g2) (g2)--(g3) (g3)--(g4) (g4)--(g5) (g5)--(g6)  (g6)--(g7) (g7)--(g8) (g8)--(g9) (g9)--(g10) (g10)--(g11) ;
	\end{tikzpicture}  
	\caption{Quiver for the theory $T^{[4^2,2^2]}(USp(12))$ with Coulomb branch $\mathcal C=\bar{\mathcal O}_{[5,3^2,1^2]}\subset \mathfrak{so}(13)$.}
	\label{fig:o13}
\end{figure}

The first nilpotent orbit with a Lusztig's Canonical Quotient of the form $\AO=\mathbb Z_2^2$ is the orbit $\mathcal O_\lambda \subset \mathfrak{so}(13)$ with  $\lambda=[5,3^2,1^2]$. The quiver with Coulomb branch $\mathcal C=\bar{\mathcal O}_\lambda$ is different from all the ones computed so far since it corresponds to the choice of \emph{two} orthogonal gauge group factors (all previous results in section \ref{sec:quivers} have either \emph{zero} or \emph{one} orthogonal gauge group factor). It is depicted in figure \ref{fig:o13} and it corresponds to theory $T^{[4^2,2^2]}(USp(12))$. 

The computed Hilbert series is:
\begin{align}
	H(t)=1+78 t + 3080 t^2 + 82082 t^3+1660658 t^4 + 27202734 t^5 + O\left(t^6\right)
\end{align}

Its plethystic logarithm:
\begin{align}
	PL(H(t))=78 t-t^2-t^4-364 t^5+O\left(t^6\right)
\end{align}


\section{Analysis of the results}\label{sec:analysis}

Analyzing the obtained results we are able to produce a prescription that fully determines the orthosymplectic quiver corresponding to $T^{\sigma}(SO(n)^\vee)$ theory such that its Coulomb branch is $\mathcal C=\bar {\mathcal O}_\lambda \subset \mathfrak{so}(n)$ with $\lambda=d_{BV}(\sigma)$. 

In order to present the prescription let us define a \emph{chain} $\Xi_i$ of $SO/O(N_j)$ groups as a subset of the quiver constituted by consecutive $SO/O(N_j)$ gauge nodes (starting with $SO/O(N_i)$, following with $SO/O(N_{i+2})$, etc. and ending with $SO/O(N_{i+2m-2})$, where the \emph{length} or number of nodes in the chain is $m$) such that:
\begin{itemize}
	\item $N_{i+2p}$ is even for $0\leq p\leq m-1$.
	\item If the quiver has gauge nodes $USp(N_{i-1})$ and/or $USp(N_{i+2m-1})$, they have flavor nodes attached to them.
	\item Gauge nodes $USp(N_{i+2p+1})$ with  $0\leq p\leq m-2$ do not have flavor node attached to them, i.e.
	\begin{align}
		\begin{aligned}
			M_{i+2p+1}&=\sigma^T_{i+2p+1}-\sigma^T_{i+2p+2}= 0,\ \ \ 0\leq p\leq m-2
		\end{aligned}
	\end{align} 
\end{itemize}

When faced with the choice $SO/O(N_j)$: The maximal magnetic lattice is the lattice for the gauge group where all \emph{chains} $\Xi_i$ are chosen to have $SO(N_{i+2p})$ as their groups for all $p$. Let us define this as the \emph{default} lattice. This magnetic lattice is chosen when $\bar A (\mathcal O_\lambda)$ is the trivial group.

When the Lusztig's Canonical Quotient is $\AO=\mathbb Z_2$ the chain $\Xi_i$ with the lowest $i$ changes. If the chain has length one, i.e. it constitutes a single node $SO(N_i)$, this node is changed to $O(N_i)$. From the point of view of the magnetic lattice this translates into projecting the magnetic lattice of $SO(N_i)$ by a $\mathbb Z_2$ action. If the chain has length $m$ bigger than one, the magnetic lattice of the groups $SO(N_i)\times SO(N_{i+2})\times \dots \times SO(N_{i+2m-2})$ needs to be projected by the diagonal $\mathbb Z_2$ acting diagonally on the individual lattices of all the groups $SO(N_{i+2p})$. This is the case of the quiver marked with the symbol ($\dagger$) in table \ref{table8}. How to perform this diagonal $\mathbb Z_2$ action is a new result that has never been discussed before as far as the authors are aware. This process is explained in detail in the next section.

When the Lusztig's Canonical Group has the form $\AO=\mathbb Z_2^k$ the $k$ chains $\Xi_i$ with lowest $i$ are changed in the same fashion\footnote{We have observed that in the case of nilpotent orbits of $\mathfrak{so} (2n+1)$ the total number of chains $\Xi_i$ coincides with $k+1$. For $\mathfrak {so} (2n)$ the total number of chains is either equal or greater than $k+1$.}: in each chain $\Xi_i$ all the lattices of the groups $SO(N_{i+2p})$ are projected by a diagonal $\mathbb Z_2$ action. This is for example the case of the quiver in figure \ref{fig:o13}: $\AO=\mathbb Z_2^2$, the first chain $\Xi_1$ has length one, the magnetic lattice of its gauge group $SO(4)$ is acted by a single $\mathbb Z_2$, so the group changes to $O(4)$; the second chain $\Xi_3$ has also length one, its gauge group $SO(8)$ changes to $O(8)$, i.e. its magnetic lattice is halved by a different $\mathbb Z_2$ action. 

\section{Diagonal $\mathbb Z_2$ actions on different magnetic lattices}\label{sec:Z2actions}

\subsection{Diagonal action on the magnetic lattices of $SO(2)\times SO(4)$}

For clarity let us discuss the quiver with $\mathcal C=\bar{\mathcal O}_{[3,2^2,1^2]}\subset \mathfrak {so}(9)$ depicted in table \ref{table8}. There are two \emph{chains}: $\Xi_1$ with groups $SO(N_1=2)$ and $SO(N_3=4)$, and $\Xi_5$ with groups $SO(N_5=4)$ and $SO(N_7=2)$. The aim is to divide the magnetic lattice of the groups in $\Xi_1$ by a single $\mathbb{Z}_2$ action.

Let the weights and dressing factors of the magnetic lattices of gauge group factors $SO(2)$, $O(2)$, $SO(4)$ and $O(4)$, as they were discussed in \cite{CHMZ14}, be described by tables \ref{tab:so2}, \ref{tab:o2}, \ref{tab:so4} and \ref{tab:o4}. Note that for the groups $O(2)$ and $O(4)$ some magnetic fluxes are defined to have vanishing dressing factor, this is equivalent to removing them from the lattice. The choice to include them in the tables becomes justified in the following discussion.

A general description of the magnetic lattices and dressing factors of gauge groups $SO(N)$ and $USp(2N)$ can be found in appendix \emph{A. Classical Casimir contribution for classical groups} of \cite{CHZ13}. For the group $SO(2N)$ the magnetic weights are the weights of the GNO dual group, $SO(2N)$. The dominant Weyl chamber of the group can be parametrized by an integer vector:
\begin{align}\label{eq:vector}
	\m=(m_1,m_2,\dots ,m_N)
\end{align}
with $m_1\geq m_2\geq \dots \geq |m_N|$. For the gauge group $SO(2N+1)$ the magnetic weights are summed over the dominant Weyl chamber in the weight lattice of the GNO dual, $USp(2N)$. These can be parametrized by the same integer vector $\m$ in equation (\ref{eq:vector}) but this time with $m_1\geq m_2\geq \dots \geq m_N$. For the gauge group $O(2N)$ the magnetic weights and dressing factors can be found in appendix \emph{A. Monopole formula for orthogonal and special orthogonal
gauge groups} of \cite{CHMZ14}. Starting with the dominant Weyl chamber of $SO(2N)$ described by (\ref{eq:vector}) the parity symmetry $\mathbb Z_2$ identifies $(m_1,m_2,\dots ,m_N)$ with $(m_1,m_2,\dots ,-m_N)$. This results in a Weyl chamber isomorphic to the Weyl chamber of $USp(2N)$ that is employed for the $SO(2N+1)$ gauge group. As explained in \cite{CHMZ14}, not only the Weyl chambers of magnetic weights of $O(2N)$ and $SO(2N+1)$ are isomorphic, but their dressing factors are also identical.

\begin{table}[t]
	\centering
	\begin{tabular}{|c|c|}
	\hline
	Monopole fluxes $\m_1$ &  $P_{SO(2)}(\m_1;t)$\\ \hline
	$(0)$ & $\frac{1}{1-t}$ \\ \hline
	$(\pm 1)$ & $\frac{1}{1-t}$ \\ \hline
	\end{tabular}
	\caption{Data for $SO(2)$ group. $\m_1$ runs from $-\infty$ to $+\infty$.}
	\label{tab:so2}
\end{table}

\begin{table}[t]
	\centering
	\begin{tabular}{|c|c|}
	\hline
	Monopole fluxes $\m_1$ &  $P_{O(2)}(\m_1;t)$\\ \hline
	$(0)$ & $\frac{1}{1-t^2}$ \\ \hline
	$(+1)$ & $\frac{1}{1-t}$ \\ \hline
	$(-1)$ & $0$ \\ \hline
	\end{tabular}
	\caption{Data for $O(2)$ group. $\m_1$ runs from $-\infty$ to $+\infty$.}
	\label{tab:o2}
\end{table}

\begin{table}[t]
	\centering
	\begin{tabular}{|c|c|}
	\hline
	Monopole fluxes $\m_2$ &  $P_{SO(4)}(\m_2;t)$\\ \hline
	$(0,0)$ & $\frac{1}{(1-t^2)^2}$ \\ \hline
	$(1,\pm1)$ & $\frac{1}{(1-t)(1-t^2)}$ \\ \hline
	$(1,0)$ & $\frac{1}{(1-t)^2}$ \\ \hline
	$(2,\pm 1)$ & $\frac{1}{(1-t)^2}$ \\ \hline
	\end{tabular}
	\caption{Data for $SO(4)$ group. $\m_2=(m_{2,1},m_{2,2})$ with $+\infty > m_{2,1}\geq |m_{2,2}|\geq 0$.}
	\label{tab:so4}
\end{table}

\begin{table}[t]
	\centering
	\begin{tabular}{|c|c|}
	\hline
	Monopole fluxes $\m_2$ &  $P_{O(4)}(\m_2;t)$\\ \hline
	$(0,0)$ & $\frac{1}{(1-t^2)(1-t^4)}$ \\ \hline
	$(1,+1)$ & $\frac{1}{(1-t)(1-t^2)}$ \\ \hline
	$(1,-1)$ & $0$ \\ \hline
	$(1,0)$ & $\frac{1}{(1-t)(1-t^2)}$ \\ \hline
	$(2,+1)$ & $\frac{1}{(1-t)^2}$ \\ \hline
	$(2,-1)$ & $0$ \\ \hline
	\end{tabular}
	\caption{Data for $O(4)$ group. $\m_2=(m_{2,1},m_{2,2})$ with $+\infty> m_{2,1}\geq |m_{2,2}|\geq 0$.}
	\label{tab:o4}
\end{table}

The combination of the magnetic lattices of gauge groups $SO(2)$, $O(2)$, $SO(4)$ and $O(4)$ gives four possibilities for chain $\Xi_1$ (the aim of this section is to present a fifth option). Let
\begin{align}
	\Gamma_i:=\Gamma^*_{\G_i^\vee}/\mathcal{W}_{\G_i^\vee}
\end{align}
where $\Gamma^*_{\G_i^\vee}$ is the weight lattice of $\G_i^\vee$, the GNO dual group of $\G_i$, and $\mathcal W_{\G_i^\vee}$ is the Weyl group of $\G_i^\vee$.

The four options are:
\begin{align}
	\begin{aligned}
		\G_1&=SO(2)\times SO(4)\\
		\G_2&=O(2)\times SO(4)\\
		\G_3&=SO(2)\times O(4)\\
		\G_4&=O(2)\times O(4)
	\end{aligned}
\end{align}

Note that the dressing factors can be obtained with usual multiplication:
\begin{align}\label{eq:product}
	\begin{aligned}
		P_1(\m_1,\m_2;t)&=P_{SO(2)}(\m_1;t)P_{SO(4)}(\m_2;t)\\
		P_2(\m_1,\m_2;t)&=P_{O(2)}(\m_1;t)P_{SO(4)}(\m_2;t)\\
		P_3(\m_1,\m_2;t)&=P_{SO(2)}(\m_1;t)P_{O(4)}(\m_2;t)\\
		P_4(\m_1,\m_2;t)&=P_{O(2)}(\m_1;t)P_{O(4)}(\m_2;t)\\
	\end{aligned}
\end{align}
	
$\Gamma_1$ is the maximal lattice of the set, and the remaining ones can be constructed via a projection by the action of a finite group:
\begin{align}
	\begin{aligned}
		\Gamma_1 &\text{ is the maximal lattice}\\
		\Gamma_2 &=\Gamma_1/\mathbb Z_2\\
		\Gamma_3 &=\Gamma_1/\mathbb Z_2\\
		\Gamma_4 &=\Gamma_1/(\mathbb Z_2\times \mathbb Z_2)\\
	\end{aligned}
\end{align}

The $\mathbb Z _2\times \mathbb Z _2$ action on $\Gamma_1$ defines three non equivalent $\mathbb Z _2$ actions on $\Gamma_1$. Let $\mathbb{Z}_2^2$ have elements $\{e,a,b,ab\}$ where $e$ is the identity, $a^2=b^2=e$ and $ab=ba$. Let the $\mathbb Z_2$ generated by $a$ be the one that gives rise to $\Gamma _2$ and the $\mathbb Z _2$ generated by $b$ the one that gives rise to $\Gamma_3$. There is a lattice that has not been considered yet, whose dressing factor cannot be obtained as a product of the form of equation (\ref{eq:product}): the projection of $\Gamma_1$ by the action of the $\mathbb Z_2$ group generated by $ab$. Let us denote this lattice as $\Gamma_5$:
\begin{align}
	\begin{aligned}
		\Gamma_5 &=\Gamma_1/\mathbb Z_2\\
	\end{aligned}
\end{align}

In order to find the lattice $\Gamma_5$ it is necessary to find an implementation of the $\mathbb Z_2$ actions generated by $a$ and $b$ that can be extended to obtain the action generated by $ab$. There are two objects that change after the projection:
\begin{itemize}
	\item The elements $\m\in \Gamma$.
	\item The dressing factor $P(\m;t)$.
\end{itemize}

Let $P_0(\m;t)$ be the dressing factor of the initial lattice $\Gamma_0$ before the projection and $P_f(\m;t)$ be the dressing factor of the resulting lattice $\Gamma_f=\Gamma_0/\gamma$, where $\gamma$ is a finite symmetry group of $\Gamma_0$. In general $\Gamma_f$ is a sublattice of $\Gamma_0$. However, one can always extend $\Gamma_f$ to $\Gamma_0$ by assigning dressing factors with zero value to the points in $\Gamma_0$ which are not in $\Gamma_f$. After this it is only necessary to describe the action of the projection on $P_0(\m;t)$ and $P_f(\m;t)$. Let us define the relation between both dressing factors with a sum over the elements in $\gamma$:
\begin{align}\label{eq:sum}
	P_f(\m;t):=\frac{1}{|\gamma|}\sum_{g\in \gamma}g\left[P_0(\m;t)\right]
\end{align}

The order of $\gamma$ is denoted by $|\gamma|$. The action of an element of the group $g\in \gamma$ on the original lattice $\Gamma_0$ is denoted via $g\left[P_0(\m;t)\right]$ which corresponds to a new dressing factor.

\begin{table}[t]
	\centering
	\begin{tabular}{|c|c|}
	\hline
	Monopole fluxes $\m_1$ & $a\left[P_{SO(2)}(\m_1;t)\right]$ \\ \hline
	$(0)$ & $\frac{1}{1+t}$ \\ \hline
	$(1)$ & $\frac{1}{1-t}$ \\ \hline
	$(-1)$ & $-\frac{1}{1-t}$ \\ \hline
	\end{tabular}
	\caption{Data for $a\left[P_{SO(2)}(\m_1;t)\right]$. $\m_1$ runs from $-\infty$ to $+\infty$.}
	\label{tab:aso2}
\end{table}

\begin{table}[t]
	\centering
	\begin{tabular}{|c|c|}
	\hline
	Monopole fluxes $\m_2$ & $b\left[P_{SO(4)}(\m_2;t)\right]$\\ \hline
	$(0,0)$ & $\frac{1}{(1-t^4)}$ \\ \hline
	$(1,+1)$ & $\frac{1}{(1-t)(1-t^2)}$ \\ \hline
	$(1,-1)$ & $-\frac{1}{(1-t)(1-t^2)}$ \\ \hline
	$(1,0)$ & $\frac{1}{(1-t^2)}$ \\ \hline
	$(2,+1)$ & $\frac{1}{(1-t)^2}$ \\ \hline
	$(2,-1)$ & $-\frac{1}{(1-t)^2}$ \\ \hline
	\end{tabular}
	\caption{Data for $b\left[P_{SO(4)}(\m_2;t)\right]$. $\m_2=(m_{2,1},m_{2,2})$ with $+\infty > m_{2,1}\geq |m_{2,2}|\geq 0$.}
	\label{tab:bso4}
\end{table}

First of all, let us see what can be learned from recasting the results of tables \ref{tab:so2}, \ref{tab:o2}, \ref{tab:so4} and \ref{tab:o4} into sums of the form of equation (\ref{eq:sum}). Let $\m_1$ label the weights in the magnetic lattice of $SO(2)$ and also $O(2)$. Let the reflexion symmetry of the magnetic lattice of $SO(2)$ be $\mathbb Z_2=\{e,a\}$ with $a^2=e$. Then we can write:
\begin{align}
	P_{O(2)}(\m_1;t)=\frac{1}{2}\left(e\left[P_{SO(2)}(\m_1;t)\right]+a\left[P_{SO(2)}(\m_1;t)\right]\right)
\end{align}

The action of the identity element leaves the dressing factor unchanged:
\begin{align}
	e\left[P_{SO(2)}(\m_1;t)\right]=P_{SO(2)}(\m_1;t)
\end{align}

Since $P_{SO(2)}(\m_1;t)$ and $P_{O(2)}(\m_1;t)$ are already known in this case, they can be used in order to obtain the action of the generator $a\in \mathbb Z_2$ on the dressing factor:
\begin{align}\label{eq:a}
	a\left[P_{SO(2)}(\m_1;t)\right]=2P_{O(2)}(\m_1;t)-P_{SO(2)}(\m_1;t)
\end{align}

Similarly for $SO(4)$ and $O(4)$, let the discrete symmetry group be $\mathbb Z_2=\{e,b\}$ with $b^2=e$ and the magnetic weights be labeled by $\m_2$. One can write the sum as:
\begin{align}
	P_{O(4)}(\m_2;t)=\frac{1}{2}\left(e\left[P_{SO(4)}(\m_2;t)\right]+b\left[P_{SO(4)}(\m_2;t)\right]\right)
\end{align}

Hence:
\begin{align}\label{eq:b}
	b\left[P_{SO(4)}(\m_2;t)\right]=2P_{O(4)}(\m_2;t)-P_{SO(4)}(\m_2;t)
\end{align}

The precise values of the \emph{dressing factors} described by equations (\ref{eq:a}) and (\ref{eq:b}) are presented in tables \ref{tab:aso2} and \ref{tab:bso4}.

Now the dressing factors of lattices $\Gamma_1$, $\Gamma_2$, $\Gamma_3$ and $\Gamma_4$ can also be rewritten as sums:
 \begin{align}\label{eq:1234}
	\begin{aligned}
		P_1(\m_1,\m_2;t)&=P_{SO(2)}(\m_1;t) P_{SO(4)}(\m_2;t)\\
		P_2(\m_1,\m_2;t)&=\frac{1}{2}\left(P_{SO(2)}(\m_1;t) P_{SO(4)}(\m_2;t)+a\left[P_{SO(2)}(\m_1;t)\right] P_{SO(4)}(\m_2;t)\right)\\
		P_3(\m_1,\m_2;t)&=\frac{1}{2}\left(P_{SO(2)}(\m_1;t) P_{SO(4)}(\m_2;t)+P_{SO(2)}(\m_1;t) b\left[P_{SO(4)}(\m_2;t)\right]\right)\\
		P_4(\m_1,\m_2;t)&=\frac{1}{4}\left(P_{SO(2)}(\m_1;t) P_{SO(4)}(\m_2;t)+a\left[P_{SO(2)}(\m_1;t)\right] P_{SO(4)}(\m_2;t)\right.\\
		&\ \ \left. +P_{SO(2)}(\m_1;t) b\left[P_{SO(4)}(\m_2;t)\right]+a\left[P_{SO(2)}(\m_1;t)\right] b\left[P_{SO(4)}(\m_2;t)\right]\right)\\
	\end{aligned}
\end{align}

\begin{table}[t]
	\centering
	\begin{tabular}{|c|c|}
	\hline
	Monopole fluxes $(\m_1,\m_2)$ & $P_5(\m_1,\m_2;t)$\\ \hline
	$(+1,0,0)$ & $\frac{1}{(1-t)(1-t^2)(1-t^4)}$ \\ \hline
	$(+1,1,+1)$ & $\frac{1}{(1-t)^2(1-t^2)}$ \\ \hline
	$(+1,1,-1)$ & $0$ \\ \hline
	$(+1,1,0)$ & $\frac{1}{(1-t)^2(1-t^2)}$ \\ \hline
	$(+1,2,+1)$ & $\frac{1}{(1-t)^3}$ \\ \hline
	$(+1,2,-1)$ & $0$ \\ \hline
	$(0,0,0)$ & $\frac{1-t+t^2}{(1-t)(1-t^2)(1-t^4)}$ \\ \hline
	$(0,1,+1)$ & $\frac{1}{(1-t)(1-t^2)^2}$ \\ \hline
	$(0,1,-1)$ & $\frac{t}{(1-t)(1-t^2)^2}$ \\ \hline
	$(0,1,0)$ & $\frac{1+t^2}{(1-t)(1-t^2)^2}$ \\ \hline
	$(0,2,+1)$ & $\frac{1}{(1-t)^2(1-t^2)}$ \\ \hline
	$(0,2,-1)$ & $\frac{t}{(1-t)^2(1-t^2)}$ \\ \hline
	$(-1,0,0)$ & $\frac{t^2}{(1-t)(1-t^2)(1-t^4)}$ \\ \hline
	$(-1,1,+1)$ & $0$ \\ \hline
	$(-1,1,-1)$ & $\frac{1}{(1-t)^2(1-t^2)}$ \\ \hline
	$(-1,1,0)$ & $\frac{t}{(1-t)^2(1-t^2)}$ \\ \hline
	$(-1,2,+1)$ & $0$ \\ \hline
	$(-1,2,-1)$ & $\frac{1}{(1-t)^3}$ \\ \hline
	\end{tabular}
	\caption{Data for $\Gamma_5$. $\m_1$ runs from $-\infty$ to $+\infty$ and $\m_2=(m_{2,1},m_{2,2})$ with $+\infty > m_{2,1}\geq |m_{2,2}|\geq 0$. The weights $(\m_1,\m_2)$ where the dressing factor vanishes correspond with points in the lattice that are removed by the projection.}
	\label{tab:gamma5}
\end{table}

After writing the dressing factors in the form (\ref{eq:1234}) it is clear that there is a sum missing from this set: the sum over only the elements $\mathbb Z_2=\{e,ab\}\subset \mathbb Z_2^2$:
\begin{align}\label{eq:gamma5}
	\begin{aligned}
		P_5(\m_1,\m_2;t)=&\frac{1}{|\mathbb Z_2|}\sum_{g\in \{e,ab\}}g\left[P_{SO(2)}(\m_1;t) P_{SO(4)}(\m_2;t)\right]\\
		=&\frac{1}{2}\left(P_{SO(2)}(\m_1;t) P_{SO(4)}(\m_2;t)+a\left[P_{SO(2)}(\m_1;t)\right] b\left[P_{SO(4)}(\m_2;t)\right]\right)\\
	\end{aligned}
\end{align}

This defines lattice $\Gamma_5=\Gamma_1/\{e,ab\}$. The specific values of the dressing factor obtained with equation (\ref{eq:gamma5}) for this lattice are given in table \ref{tab:gamma5}. We draw the quiver related to this lattice as the quiver in table \ref{table8} labeled with partition $\lambda=[3,2^2,1^2]$. The brackets in the quiver denote that the gauge groups $SO(2)$ and $SO(4)$ in the first chain $\Xi_1$ contribute not with the entire magnetic lattice but with a $\mathbb Z_2$ diagonal projection of the same. When this quiver is used on the computation of the Coulomb branch of the corresponding $T^{[4^2]}(SO(9)^\vee)$ quiver the Hilbert series obtained with the \emph{monopole formula} is indeed the Hilbert series of the closure of nilpotent orbit $\bar{\mathcal O}_{[3,2^2,1^2]}\subset \mathfrak {so}(9)$, presented in the $9^{th}$ row of table \ref{tab:HS9}.

\begin{figure}[t]
	\centering
\begin{tikzpicture}
				\tikzstyle{gauge1} = [draw=none,minimum size=0.35cm,fill=blue,circle, draw];
				\tikzstyle{gauge2} = [draw=none,minimum size=0.35cm,fill=red,circle, draw];
				\tikzstyle{flavour1} = [draw=none,minimum size=0.35cm,fill=blue, regular polygon,regular polygon sides=4,draw];
				\tikzstyle{flavour2} = [draw=none,minimum size=0.35cm,fill=red, regular polygon,regular polygon sides=4,draw];
				\node (g1) [gauge2,label=below:{$2$}] {};
				\node (g2) [gauge1,right of=g1,label=below:{$2$}] {};
				\node (f2) [flavour2,above of=g4,label=above:{$1$}] {};
				\node (g3) [gauge2,right of=g2,label=below:{$4$}] {};
				\node (g4) [gauge1,right of=g3,label=below:{$4$}] {};
				\node (g5) [gauge2,right of=g4,label=below:{$5$}] {};
				\node (g6) [gauge1,right of=g5,label=below:{$4$}] {};
				\node (f6) [flavour2,above of=g6,label=above:{$1$}] {};
				\node (g7) [gauge2,right of=g6,label=below:{$[4$}] {};
				\node (g8) [gauge1,right of=g7,label=below:{$2$}] {};
                \node (g9) [gauge2,right of=g8,label=below:{$2]$}] {};
				\draw (g4)--(f2) (g6)--(f6)
				(g1)--(g2) (g2)--(g3) (g3)--(g4) (g4)--(g5) (g5)--(g6) (g6)--(g7) (g7)--(g8) (g8)--(g9);
				\end{tikzpicture}                     
	\caption{Quiver for the theory $T^{[6,4]}(USp(10))$ with Coulomb branch $\mathcal C=\bar{\mathcal O}_{[3,2^2,1^4]}\subset \mathfrak{so}(11)$.}
	\label{so11}
\end{figure}
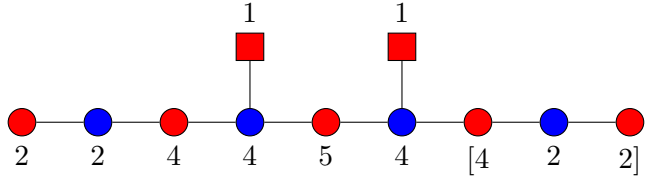

\begin{table}[t]

	\centering
	\begin{tabular}{|c|c|c|}
		\hline
		Partition ($\lambda$)   &Hilbert series  &Plethystic logarithm\\ \hline

		\raisebox{-.8\height}{$[3,2^2,1^4]$}	               & 

\raisebox{-.8\height}{
		$ \dfrac{
	 \splitfrac{1+25t+326t^2+2433t^3+10582t^4+26423t^5+36250t^6 }
	 {+26423t^7+10582t^8+2433t^9+326t^{10}+25t^{11}+t^{12}}
	 }{(1 - t)^{28} (1 + t)^{-2}}$ 	}

		 & \parbox[t]{3.5cm}{$55t-t^2-517t^3+2706t^4+5610t^5+O(t^6)$}

			\\ \hline

	\end{tabular}
	\caption{Hilbert series and plethystic logarithm for the Coulomb branch of quiver in Figure \ref{so11}.}
	\label{so112}
\end{table}

\paragraph{Second example.} The $\Gamma_5$ lattice can also be used now for partition $\lambda=[3,2^2,1^4]$, the next case that requires a diagonal $\mathbb Z_2$ action on $SO(2)$ and $SO(4)$. The theory is $T^{[6,4]}(SO(11)^\vee)$, its quiver is given in figure \ref{so11} and its Hilbert series and plethystic logarithm are given in Table \ref{so112}. The Coulomb branch of this quiver is found to be the closure of nilpotent orbit $\mathcal{O}_{[3,2^2,1^4]}\subset \mathfrak{so}(11)$.

\subsection{The general case}

In general we have the diagonal $\mathbb Z_2=\{e,a_1a_2\dots a_m\}$ where $\mathbb Z_2\subset \mathbb Z_2^m$ and $\mathbb Z_2^m$ is generated by $a_k$ with $k=1,2,\dots,m$ acting on a chain $\Xi_i$ of length $m$ with groups $SO(N_{i+2p})$ and $p=0,1,\dots ,m-1$. The magnetic lattice of the gauge group $\G=\times_p SO(N_{i+2p})$, projected by the $\mathbb Z_2$ action, is defined by the dressing factor:
\begin{align}
	\begin{aligned}
		P_{\left(\Xi_i\right)}(\m_1,\dots,\m_m;t)=\frac{1}{2}\left(\prod_{k=1}^{m}P_{SO(N_{i+2k-2})}(\m_k;t)+\prod_{k=1}^{m}a_k\left[P_{SO(N_{i+2k-2})}(\m_k;t)\right]\right)
	\end{aligned}
\end{align}

where $(\m_1,\dots, \m_m)$ run over the magnetic weights in the lattice $\Gamma^*_{\G^\vee}/\mathcal W_{\G^\vee}$ before the projection. $a_k\left[P_{SO(N_{i+2k-2})}(\m_k;t)\right]$ is defined in terms of the $SO(N_{i+2k-2})$ and the $O(N_{i+2k-2})$ dressing factors:
\begin{align}
	\begin{aligned}
		a_k\left[P_{SO(N_{i+2k-2})}(\m_k;t)\right]&:=2P_{O(N_{i+2k-2})}(\m_k;t)-P_{SO(N_{i+2k-2})}(\m_k;t)
	\end{aligned}
\end{align}
where $P_{O(N_{i+2k-2})}(\m_k;t)$ is extended over the weights that are in the magnetic lattice of $SO(N_{i+2k-2})$ but not in the magnetic lattice of $O(N_{i+2k-2})$ by defining it to be zero at those weights.

\section{Conclusion and outlook}

The result of this project is an answer to the question \emph{Are the gauge symmetry factors $O(2N_i)$ or $SO(2N_i)$ on the $3d\ \mathcal N=4$ orthosymplectic quiver of $T^\sigma (SO(n)^\vee)$ theories?} The answer is given by the Lusztig's Canonical Quotient $\AO$ where $\lambda = d_{VB}(\sigma)$. If $\AO$ is the trivial group all gauge symmetry factors are $SO(2N_i)$. For a group of the form $\AO=\mathbb Z_2^k$ the magnetic lattice of the $SO(2N_i)$ gauge factors is projected by $k$ different $\mathbb{Z}_2$ actions, in the way described in section \ref{sec:analysis}.

Sometimes a single $\mathbb Z_2$ group acts diagonally on multiple $SO(2N_i)$ magnetic lattices. The way of implementing this action is developed here for the first time, section \ref{sec:Z2actions}. This process widens the current reach of the \emph{monopole formula} and opens the door to many new exciting computations in the field of $3d\ \mathcal{N}=4$ gauge theories.

The results of this note can be complemented by analyzing the nature of the Type IIB brane configurations that give rise to the orthosymplectic quivers discussed. This analysis is currently under development and can show different brane configurations corresponding to different choices of $SO/O(N_i)$ gauge group factors.

So far, all the moduli spaces that have been discussed here are closures of special nilpotent orbits. An interesting direction in which this project could be further developed is the analysis of quivers with different choices of $SO/O(N_i)$ gauge factors such that their Coulomb branches are no longer closures of nilpotent orbits. The study of the geometry and the physics behind these moduli spaces is a very rich and challenging problem that remains to be addressed.

\section*{Acknowledgements}

We would like to thank Gong Cheng, Lingtong Chen, Giulia Ferlito, Yabo Li, Rudolph Kalveks, Marcus Sperling and Yidi Zhao  for helpful conversations during the development of this project. S.C. is supported  by an EPSRC DTP studentship EP/M507878/1. A.H. is supported by STFC Consolidated Grant ST/J0003533/1, and EPSRC Programme Grant EP/K034456/1.

\bibliography{main}
\bibliographystyle{JHEP}
\end{document}